\documentclass[12pt,letterpaper]{article}
\usepackage{epsfig,rotating,setspace,latexsym,amsmath,epsf,amssymb,bm}
\usepackage{cite}

\title{A New Data Processing Inequality and Its Applications in Distributed Source and Channel Coding\thanks{This work was supported by NSF Grants CCR $03$-$11311$, CCF $04$-$47613$ and CCF $05$-$14846$. It was presented in part at the Asilomar Conference on Signals, Systems and Computers, Pacific Grove, CA, October 2005 \cite{Kang:2005}, the Conference on Information Sciences and Systems (CISS), Princeton, NJ, March 2006 \cite{Kang:2006a}, and the IEEE International Symposium on Information Theory (ISIT), Seattle, WA, July 2006 \cite{Kang:2006b}.}}

\author{Wei Kang \qquad Sennur Ulukus \\
\normalsize Department of Electrical and Computer Engineering\\
\normalsize University of Maryland, College Park, MD 20742 \\
\normalsize {\it wkang@eng.umd.edu} \qquad {\it ulukus@umd.edu}
}

\newtheorem{Theo}{Theorem}
\newtheorem{Lem}{Lemma}

\newtheorem{Def}{Definition}
\newenvironment{proof}[1]{\medskip\par\noindent
{\bf Proof:\,}\,#1}{{\mbox{\,$\blacksquare$}\par}}
\begin{document}
\maketitle \vspace*{-0.5cm}
\begin{center}
{\bf Abstract}
\end{center}

\textrm{In the distributed coding of correlated sources, the problem of characterizing the joint probability distribution of a pair of random variables satisfying an $n$-letter Markov chain arises. The exact solution of this problem is intractable. In this paper, we seek a single-letter necessary condition for this $n$-letter Markov chain.
To this end, we propose a new data processing inequality on a new measure of correlation by means of spectrum analysis. Based on this new data processing inequality, we provide a single-letter necessary condition for the required joint probability distribution. 
We apply our results to two specific examples involving the distributed coding of correlated sources: multi-terminal rate-distortion region and multiple access channel with correlated sources, and
propose new necessary conditions for these two problems.}
\clearpage
\section{Problem Formulation}\label{introPF}
In this paper, we consider a pair of correlated discrete source sequences with length $n$, $(U^n, V^n) = \{(U_1,V_1),\dots,(U_n, V_n)\}$, which are  independent and identically distributed (i.i.d.) in time, i.e.,
\begin{equation}
p(u^n,v^n)=\prod_{i=1}^{n}p(u_i,v_i)
\end{equation}
and
\begin{equation}
p(u_i,v_i)=p(u,v),\qquad  i=1,\dots,n
\end{equation}
where the single-letter joint distribution $p(u,v)$ is defined on the alphabet $\mathcal{U}\times\mathcal{V}$. 
Let $(X_1, X_2)$ be two random variables such that $(X_1,X_2, U^n, V^n)$ satisfies
\begin{equation}
p(x_1,x_2,u^n,v^n)=p(u^n,v^n)p(x_1|u^n)p(x_2|v^n)
\end{equation}
or equivalently\footnote{$X_1=f_1(U^n)$ and $X_2=f_2(V^n)$ is a degenerate case.}, 
\begin{equation}
X_{1}\longrightarrow U^n\longrightarrow V^n\longrightarrow X_{2}\nonumber
\end{equation}
This Markov chain appears in some problems involving the distributed coding of correlated sources. For example, in distributed rate-distortion problem \cite{Tung:1978,Housewright:1977, Berger:1978}, $(X_{1}, X_{2})$ is used to reconstruct, $(\hat{U}^n,\hat{V}^n)$,  an estimate of the sources $(U^n, V^n)$, and in the problem of multiple access channel with correlated sources \cite{Cover:1980, Ahlswede:1983}, $(X_{1}, X_{2})$ is sent though a multiple access channel in one channel use. 
Although these specific problems have been studied separately in their own contexts, the common nature of these problems, the distributed coding of correlated sources, enables us to conduct a general study, which will be applicable to these specific problems.

The study of the converse proofs of (or the necessary conditions for) the above specific problems raises the following questions. 
We know that the correlation between $(X_1, X_2)$ is limited, if a single-letter Markov chain $X_{1}\longrightarrow U\longrightarrow V\longrightarrow X_{2}$ is to be satisfied. With the help of more letters of the sources, i.e.,~$X_{1}\longrightarrow U^n\longrightarrow V^n\longrightarrow X_{2}$ with $n$ larger than $1$, the correlation between $(X_1, X_2)$ may increase. The question here is how correlated  $(X_1, X_2)$ can be, when $n$ goes to infinity. More specifically, can they be arbitrarily correlated? If not, then, how much extra correlation can $(X_1, X_2)$ gain when $n$ goes from $1$ to $\infty$?   To answer these questions, we need to determine the set of all ``valid" joint probability distributions $p(x_{1}, x_{2})$, if $X_{1}\longrightarrow U^n\longrightarrow V^n\longrightarrow X_{2}$ is to be satisfied with $n$ going to infinity\footnote{We are also interested in determining the set of all ``valid" probability distributions $p(x_{1},x_{2}, u_1,v_1)$, or the set of all  ``valid" probability distributions $p(x_{1},x_{2},u_1,u_2,v_1,v_2)$, etc., if this Markov chain constraint is to be satisfied.}, i.e.,
\begin{equation}\label{suff}
\mathcal{S}_{X_1X_2}\triangleq\{p(x_1,x_2):X_{1}\longrightarrow U^n\longrightarrow V^n\longrightarrow X_{2},\quad n\rightarrow \infty\}
\end{equation}

We note that it is practically impossible to exhaust the elements in the set $\mathcal{S}_{X_1X_2}$ by searching over all conditional distribution pairs $\left(p(x_1|u^n), p(x_2|v^n)\right)$ when $n\rightarrow \infty$. In other words, determining the set of all possible probability distributions $p(x_1,x_2)$ satisfying the $n$-letter Markov chain, i.e.,~the set $\mathcal{S}_{X_1X_2}$, seems computationally intractable.
To avoid this problem, we  seek a single-letter necessary condition for the above $n$-letter Markov chain. The resulting set, characterized by computable single-letter constraints, will contain the target set $\mathcal{S}_{X_1X_2}$.

The most intuitive necessary condition for a Markov chain is the data processing inequality \cite[p. 32]{Cover:1991}, i.e.,~if $X_1\longrightarrow U^n\longrightarrow V^n\longrightarrow X_2$, then
 \begin{align}
I(X_1;X_2)\le I(U^n;V^n)=nI(U;V)\label{datap}
 \end{align}
Since $I(U^n;V^n)$ increases linearly with $n$,  the constraint in (\ref{datap}) will be loose when $n$ is sufficiently large. 
%
Although the data processing inequality in its usual form
does not prove useful in this problem, we will still use the basic methodology of employing
a data processing inequality to find a necessary condition for the $n$-letter Markov chain under consideration. For this, we will introduce a new measure of correlation, and develop  a new data processing inequality based on this new measure of correlation.

Spectrum analysis has been instrumental in the study of some properties of pairs of correlated random variables,
especially, those of i.i.d. sequences of pairs of correlated random variables,
e.g.,~common information in \cite{Witsenhausen:1975}
and isomorphism in \cite{Marton:1981}.
In this paper, we use spectrum analysis to introduce a new data processing inequality,
which provides a single-letter necessary condition for the joint
distributions satisfying the $n$-letter Markov chain.

\section{Main Results}\label{math}
\subsection{Some Preliminaries}
In this section, we provide some basic results which will be used in our later development. The concepts used here are originally introduced by Witsenhausen in \cite{Witsenhausen:1975} in the context of operator theory. Here, we focus on the finite alphabet case,  and  derive our results by means of matrix theory.

We first introduce our matrix notation for probability distributions.
For a pair of discrete random variables $X$ and
$Y$, which take values in $\mathcal{X}$ and $\mathcal{Y}$, respectively, the $|\mathcal{X}|\times |\mathcal{Y}|$ joint probability distribution matrix $P_{XY}$ is defined as 
\begin{equation}
P_{XY}(i,j)\triangleq
Pr(X=x_i, Y=y_j)
\end{equation} 
where $P_{XY}(i,j)$ denotes the $(i,j)$-th element
of the matrix $P_{XY}$.  
The marginal distribution matrix of a random variable $X$, $P_X$,  is defined as a diagonal matrix with 
\begin{equation}
P_{X}(i,i)\triangleq Pr(X=x_i)
\end{equation}
and the vector-form marginal distribution, $p_X$,  is defined as\footnote{In this paper, we only consider the case where $p_X$ is a positive vector.} 
\begin{equation}
p_X(i)\triangleq Pr(X=x_i)
\end{equation}
or equivalently
$
p_X=P_X \mathbf{e}
$, 
where $\mathbf{e}$ is the vector of all ones. $p_X$ can also be  defined as $p_X\triangleq P_{XY}$ for some degenerate random variable $Y$ whose alphabet size $|\mathcal{Y}|$ is equal to one. For convenience, we define 
\begin{equation}
p_X^{\frac{1}{2}}\triangleq P_X^{\frac{1}{2}}\mathbf{e}
\end{equation}
For conditional distributions, we define matrix $P_{XY|z}$ as
\begin{equation}
P_{XY|z}(i,j)\triangleq Pr(X=x_i, Y=y_j|Z=z)
\end{equation}
The vector-form conditional distribution $p_{X|z}$ is defined as 
\begin{equation}
p_{X|z}(i)\triangleq Pr(X=x_i|Z=z)
\end{equation}
or equivalently, $p_{X|z}(i)\triangleq P_{XY|z}$ for some degenerate random variable $Y$ whose alphabet size $|\mathcal{Y}|$ is equal to one.

We define a new matrix, $\tilde{P}_{XY}$, which will
play an important role in the rest of the paper, as
\begin{equation}
\tilde{P}_{XY}\triangleq P_X^{-\frac{1}{2}}P_{XY}P_Y^{-\frac{1}{2}}\label{def}
\end{equation}
Since $p_X\triangleq P_{XY}$ for some degenerate random variable $Y$ whose alphabet size $|\mathcal{Y}|$ is equal to one, we define
\begin{equation}
\tilde{p}_{X}=P_X^{-\frac{1}{2}}P_{XY}P_Y^{-\frac{1}{2}}=P_X^{-\frac{1}{2}}p_{X}=p_X^{\frac{1}{2}}\label{def1}
\end{equation}
The counterparts for conditional distributions, $\tilde{P}_{XY|z}$ and $\tilde{p}_{X|y}$, can be defined similarly.

A valid joint distribution matrix, $P_{XY}$,  is a matrix whose entries are non-negative and sum to $1$.
Due to this constraint, not every matrix will qualify as a $\tilde{P}_{XY}$ corresponding to a joint distribution matrix as defined in (\ref{def}). A necessary and sufficient condition for $\tilde{P}_{XY}$ to correspond to a joint distribution matrix 
is given in Theorem \ref{iff} below, which
identifies the spectral properties of $\tilde{P}_{XY}$.
Before stating the theorem, we provide a lemma  and a definition regarding  stochastic matrices, which will be used in the proof of the theorem.
\begin{Def}\cite[p. 48]{Berman:1979}
A square matrix $T$ of order $n$ is called (row) stochastic if 
\begin{equation}
T(i,j)\ge 0\qquad i,j=1,\dots,n,\qquad\qquad
\sum_{j=1}^n T(i,j)=1\qquad i=1,\dots,n
\end{equation}
\end{Def}

\begin{Lem}\cite[p. 49]{Berman:1979}\label{sto}
The spectral radius of a stochastic matrix is $1$. A non-negative matrix $T$ is stochastic if and only if $\mathbf{e}$ is an eigenvector of $T$ corresponding to the eigenvalue $1$.
\end{Lem}

\begin{Theo}\label{iff}
A non-negative matrix $P$ is a joint distribution matrix with marginal distributions $P_X$ and $P_Y$, i.e.,~$P\mathbf{e}=p_X\triangleq P_X\mathbf{e}$ and $P^T\mathbf{e}=p_Y\triangleq P_Y\mathbf{e}$,
if and only if the singular value decomposition (SVD) of the non-negative matrix $\tilde{P}\triangleq P_X^{-\frac{1}{2}}PP_Y^{-\frac{1}{2}}$ satisfies 
\begin{equation}
\tilde{P}=M\Lambda N^T =p_X^{\frac{1}{2}}(p_Y^{\frac{1}{2}})^T+\sum_{i=2}^l \lambda_i \bm{\mu}_i\bm{\nu}_i^T\label{fun}
\end{equation}
where $M\triangleq [\bm{\mu}_1,\dots, \bm{\mu}_l]$ and $N\triangleq [\bm{\nu}_1,\dots, \bm{\nu}_l]$ are two unitary matrices, $\Lambda\triangleq \mathrm{diag}[\lambda_1,\dots,\lambda_l]$ and $l=\min(|\mathcal{X}|,|\mathcal{Y}|)$;
$\bm{\mu}_1=p_X^{\frac{1}{2}}$, $\bm{\nu}_1=p_Y^{\frac{1}{2}}$, and $\lambda_1=1\ge\lambda_2\ge\dots\ge\lambda_l\ge 0$. That is, all of the singular values of $\tilde{P}$ are between $0$ and $1$, the largest singular value of $\tilde{P}$ is $1$, and the corresponding left and right singular vectors are $p_X^{\frac{1}{2}}$ and $p_Y^{\frac{1}{2}}$.
\end{Theo}
\begin{proof} Let $\tilde{P}$ satisfy (\ref{fun}), then
\begin{align}
P_X^{\frac{1}{2}}\tilde{P} P_Y^{\frac{1}{2}}\mathbf{e}&=P_X^{\frac{1}{2}}
\left(p_X^{\frac{1}{2}}(p_Y^{\frac{1}{2}})^T+\sum_{i=2}^l \lambda_i \bm{\mu}_i\bm{\nu}_i^T\right)
p_Y^{\frac{1}{2}}\nonumber\\
&=P_X^{\frac{1}{2}}p_X^{\frac{1}{2}}(p_Y^{\frac{1}{2}})^T p_Y^{\frac{1}{2}}+
P_X^{\frac{1}{2}}\sum_{i=2}^l \lambda_i \bm{\mu}_i\bm{\nu}_i^T \bm{\nu}_1\nonumber\\
&=p_X
\end{align}
Similarly, $\mathbf{e}^T P_X^{\frac{1}{2}}\tilde{P} P_Y^{\frac{1}{2}}=p_Y^T$. Thus, the non-negative matrix $P_X^{\frac{1}{2}}\tilde{P} P_Y^{\frac{1}{2}}$ is a joint distribution matrix with marginal distributions $p_X$ and $p_Y$.

Conversely, we consider a joint distribution $P$ with marginal distributions $p_X$ and $p_Y$.
We need to show that the singular values of $\tilde{P}$ lie in $[0,1]$, the largest singular value is  equal to $1$, and $p_X^{\frac{1}{2}}$ and $p_Y^{\frac{1}{2}}$, respectively, are the left and right singular vectors
corresponding to the singular value $1$. To this end, we first construct a Markov chain $X\rightarrow Y\rightarrow Z$ with $P_{XY}=P_{ZY}=P$ (this construction comes from \cite{Witsenhausen:1975}). Note that this also implies $P_X=P_Z$, $\tilde{P}_{XY}=\tilde{P}_{ZY}=\tilde{P}$,  and $P_{X|Y}=P_{Z|Y}$.
The special structure of the constructed Markov chain provides the following:
\begin{align}
P_{X|Z}&=P_{X|Y}P_{Y|Z}\nonumber\\
&=P_{X|Y}P_{Y|X}\nonumber\\
&=PP_Y^{-1}P^T P_X^{-1}\nonumber\\
&=P_X^{\frac{1}{2}}(P_X^{-\frac{1}{2}}PP_Y^{-\frac{1}{2}})(P_Y^{-\frac{1}{2}}P^T P_X^{-\frac{1}{2}})
P_X^{-\frac{1}{2}}\nonumber\\
&=P_X^{\frac{1}{2}}\tilde{P}\tilde{P}^T P_X^{-\frac{1}{2}}
\end{align}
which implies that the matrix $P_{X|Z}$ is similar to the matrix $\tilde{P}\tilde{P}^T$ \cite[p. 44]{Horn:1985}.
Therefore, all the eigenvalues of $P_{X|Z}$  are the eigenvalues of $\tilde{P}\tilde{P}^T$ as well, and if 
$\bm{\nu}$ is a left eigenvector of $P_{X|Z}$ corresponding to an eigenvalue $\lambda$, then $P_X^{\frac{1}{2}}\bm{\nu}$ is a left eigenvector of $\tilde{P}\tilde{P}^T$ corresponding to the same eigenvalue.

We note that $P_{X|Z}^T$ is a stochastic matrix, therefore, from Lemma \ref{sto}, $\mathbf{e}$ is a left eigenvector of $P_{X|Z}$ corresponding the eigenvalue $1$, which is equal to the spectral radius of $P_{X|Z}$. Since $P_{X|Z}$ is similar to $\tilde{P}\tilde{P}^T$, we have that $p_{X}^{\frac{1}{2}}$ is a left eigenvector of $\tilde{P}\tilde{P}^T$ with eigenvalue $1$, 
%
and all the eigenvalues of $\tilde{P}\tilde{P}^T$ lie in $[-1,1]$. In addition, $\tilde{P}\tilde{P}^T$ is a symmetric positive semi-definite matrix, which implies that the eigenvalues of $\tilde{P}\tilde{P}^T$
are real and non-negative. 
Since the eigenvalues of $\tilde{P}\tilde{P}^T$ are non-negative, and the largest eigenvalue is equal to $1$, we conclude that all of the eigenvalues of $\tilde{P}\tilde{P}^T$ lie in  the interval $[0,1]$.

The singular values of $\tilde{P}$ are the square roots of the eigenvalues of $\tilde{P}\tilde{P}^T$, and the left singular vectors of $\tilde{P}$ are the eigenvectors of $\tilde{P}\tilde{P}^T$. Thus, the singular values of $\tilde{P}$ lie in $[0,1]$, the largest singular value is  equal to $1$, and $p_X^{\frac{1}{2}}$ is a left singular vector corresponding to the singular value $1$. The corresponding right singular vector is
\begin{align}
\bm{\nu}_1^T&=\bm{\mu}_1^T\tilde{P}=(p_X^{\frac{1}{2}})^T P_X^{-\frac{1}{2}}P P_Y^{-\frac{1}{2}}
=\mathbf{e}^T P P_Y^{-\frac{1}{2}}
=p_Y^T P_Y^{-\frac{1}{2}}
=(p_Y^{\frac{1}{2}})^T
\end{align}
which concludes the proof.
\end{proof}

This theorem implies that there is a one-to-one relationship between $P$ and $\tilde{P}$. It is easy to see from  (\ref{def}) that there is a unique $\tilde{P}$ for every $P$. Conversely, any given $\tilde{P}$ satisfying (\ref{fun}) gives a unique pair of marginal distributions $(P_X, P_Y)$, which is specified by the left and right positive singular vectors corresponding to its largest singular value\footnote{We observe that there may exist multiple singular values equal to $1$, but $\bm{\mu}_1$ and $\bm{\nu}_1$ are the only positive singular vectors.}. Then, from (\ref{def}), using $\tilde{P}$ and $(P_X, P_Y)$ given by its singular vectors, we obtain a unique $P$ as
\begin{equation}
P= P_X^{\frac{1}{2}}\tilde{P}P_Y^{\frac{1}{2}}
\end{equation}
Because of this one-to-one relationship, exploring all possible joint distribution matrices $P$ is equivalent to exploring all possible non-negative matrices $\tilde{P}$ satisfying (\ref{fun}).

Here, $\lambda_2,\dots, \lambda_l$ can be viewed as a group of quantities, which measures the correlation between random variables $X$ and $Y$. We note that when $\lambda_2=\cdots=\lambda_l=1$, $X$ and $Y$ are fully correlated, and,  when $\lambda_2=\cdots=\lambda_l=0$, $X$ and $Y$ are independent. In all the cases between these two extremes, $X$ and $Y$ are arbitrarily correlated. Moreover, Witsenhausen showed that $X$ and $Y$ have a common data if and only if $\lambda_2=1$ \cite{Witsenhausen:1975}. In the next section, we will propose a new data processing inequality with respect to these new measures of correlation, $\lambda_2,\dots,\lambda_l$. By utilizing this new data processing inequality, we will provide a single-letter necessary condition for the $n$-letter Markov chain $X_1\longrightarrow U^n
\longrightarrow V^n \longrightarrow X_2$.

\subsection{A New Data Processing Inequality}\label{dpi}
In this section, first, we introduce a new data processing inequality in the following theorem. Here, we provide a lemma that will be used in the proof of the theorem.
\begin{Lem}\label{pro}\cite[p. 178]{Horn:1991}
For matrices $A$ and $B$
\begin{equation}
\lambda_{i}(AB)\le\lambda_{i}(A)\lambda_1(B)
\end{equation}
where $\lambda_{i}(\cdot)$ denotes the $i$-th largest singular value of a matrix.
\end{Lem}
\begin{Theo}\label{sigpro}
If $X\rightarrow Y\rightarrow Z$, then
\begin{align}\label{newd}
\lambda_i(\tilde{P}_{XZ})\le\lambda_i(\tilde{P}_{XY})\lambda_2(\tilde{P}_{YZ})&\le\lambda_i(\tilde{P}_{XY})
 \end{align}
where  $i=2,\dots, \mathrm{rank}(\tilde{P}_{XZ})$.
\end{Theo}
\begin{proof}
From the structure of the Markov chain, and from the definition of $\tilde{P}_{XY}$ in (\ref{def}), we have
\begin{align}
\tilde{P}_{XZ}&=P_X^{-\frac{1}{2}}P_{XZ}P_Z^{-\frac{1}{2}}\nonumber\\&
=P_X^{-\frac{1}{2}}P_{XY}P_Y^{-\frac{1}{2}}P_Y^{-\frac{1}{2}}P_{YZ}P_Z^{-\frac{1}{2}}\nonumber\\
&=\tilde{P}_{XY}\tilde{P}_{YZ}\label{prod}
\end{align}
Using (\ref{fun}) for $\tilde{P}_{XZ}$, we obtain
\begin{align}
\tilde{P}_{XZ}=&p_X^{\frac{1}{2}}(p_Z^{\frac{1}{2}})^T+
\sum_{i=2}^l \lambda_i(\tilde{P}_{XZ}) \bm{\mu}_i(\tilde{P}_{XZ})\bm{\nu}_i(\tilde{P}_{XZ})^T\label{lef}
\end{align}
and applying (\ref{fun}) to $\tilde{P}_{XY}$ and $\tilde{P}_{YZ}$ yields
\begin{align}
\tilde{P}_{XY}&\tilde{P}_{YZ}\nonumber\\
=&\left(p_X^{\frac{1}{2}}(p_Y^{\frac{1}{2}})^T\!+\!
\sum_{i=2}^l \lambda_i(\tilde{P}_{XY}) \bm{\mu}_i(\tilde{P}_{XY})\bm{\nu}_i(\tilde{P}_{XY})^T\right)
\left(p_Y^{\frac{1}{2}}(p_Z^{\frac{1}{2}})^T+
\sum_{i=2}^l \lambda_i(\tilde{P}_{YZ}) \bm{\mu}_i(\tilde{P}_{YZ})\bm{\nu}_i(\tilde{P}_{YZ})^T\right)\nonumber\\
=&p_X^{\frac{1}{2}}(p_Z^{\frac{1}{2}})^T+
\left(\sum_{i=2}^l \lambda_i(\tilde{P}_{XY}) \bm{\mu}_i(\tilde{P}_{XY})\bm{\nu}_i(\tilde{P}_{XY})^T\right)
\left(\sum_{i=2}^l \lambda_i(\tilde{P}_{YZ}) \bm{\mu}_i(\tilde{P}_{YZ})\bm{\nu}_i(\tilde{P}_{YZ})^T\right)
\label{righ}
\end{align}
where the two cross-terms vanish because $p_Y^{\frac{1}{2}}$ plays the roles of both $\bm{\nu}_1(\tilde{P}_{XY})$ and $\bm{\mu}_1(\tilde{P}_{YZ})$, and therefore, $p_Y^{\frac{1}{2}}$ is orthogonal to both $\bm{\nu}_i(\tilde{P}_{XY})$ and $\bm{\mu}_j(\tilde{P}_{YZ})$, for all $i,j \neq 1$.
Using (\ref{prod}) and equating (\ref{lef}) and (\ref{righ}), we obtain
\begin{align}
\sum_{i=2}^l \lambda_i(\tilde{P}_{XZ})& \bm{\mu}_i(\tilde{P}_{XZ})\bm{\nu}_i(\tilde{P}_{XZ})^T\nonumber\\
=&\left(\sum_{i=2}^l \lambda_i(\tilde{P}_{XY}) \bm{\mu}_i(\tilde{P}_{XY})\bm{\nu}_i(\tilde{P}_{XY})^T\right)
 \left(\sum_{i=2}^l \lambda_i(\tilde{P}_{YZ}) \bm{\mu}_i(\tilde{P}_{YZ})\bm{\nu}_i(\tilde{P}_{YZ})^T\right)
\label{bela}
\end{align}
The proof is completed by applying Lemma \ref{pro} to (\ref{bela}) and also by noting that $\lambda_2(\tilde{P}_{YZ})\le 1$ from Theorem \ref{iff}.
\end{proof}

Theorem \ref{pro} is a new data processing inequality in the sense that the processing from $Y$ to $Z$ reduces the correlation measure $\lambda_i$, i.e.,~the correlation between $X$ and $Z$, $\lambda_{i}(\tilde{P}_{XZ})$, is less than  or equal to the correlation measure between $X$ and $Y$, $\lambda_i(\tilde{P}_{XY})$. We note that this theorem is similar to the data processing inequality in \cite[p. 32]{Cover:1991} except instead of mutual information, we use $\lambda_i(\tilde{P}_{XY})$ as the correlation measure. In the sequel, we will show that this new data processing inequality helps us develop a necessary condition for the $n$-letter Markov chain while the data processing inequality in its usual form \cite[p. 32]{Cover:1991} is not useful in this context.

\subsection{A Necessary Condition}\label{iid}
Now, we switch our attention to i.i.d. sequences of correlated sources. Let $(U^n, V^n)$ be a pair of i.i.d. (in time) sequences, where each  letter of these sequences satisfies a joint distribution $P_{UV}$. Thus, the joint distribution of the sequences is $P_{U^n V^n}=P_{UV}^{\otimes n}$, where $A^{\otimes1}\triangleq A$, $A^{\otimes k}\triangleq A\otimes A^{\otimes (k-1)}$, and $\otimes$ denotes the Kronecker product of matrices \cite{Horn:1985}.

From (\ref{def}), we know that
\begin{equation}
P_{UV}=P_U^{\frac{1}{2}}\tilde{P}_{UV}P_V^{\frac{1}{2}}
\end{equation}
Then,
\begin{equation}
P_{U^n V^n}=P_{UV}^{\otimes n}=(P_U^{\frac{1}{2}}\tilde{P}_{UV}P_V^{\frac{1}{2}})^{\otimes n}
=(P_U^{\frac{1}{2}})^{\otimes n}\tilde{P}_{UV}^{\otimes n}(P_V^{\frac{1}{2}})^{\otimes n}
\end{equation}
We also have $P_{U^n}=P_U^{\otimes n}$ and $P_{V^n}=P_V^{\otimes n}$. Thus,
\begin{align}
\tilde{P}_{U^nV^n}&\triangleq P_{U^n}^{-\frac{1}{2}} P_{U^n V^n} P_{V^n}^{-\frac{1}{2}}\nonumber\\
&=(P_U^{-\frac{1}{2}})^{\otimes n}(P_U^{\frac{1}{2}})^{\otimes n}\tilde{P}_{UV}^{\otimes n}
(P_V^{\frac{1}{2}})^{\otimes n}(P_V^{-\frac{1}{2}})^{\otimes n}\nonumber\\
&=\tilde{P}_{UV}^{\otimes n}
\end{align}
Now, applying SVD to $\tilde{P}_{U^nV^n}$, we have
\begin{equation}\label{SVDPN}
\tilde{P}_{U^nV^n}=M_n\Lambda_nN_n^T=\tilde{P}_{UV}^{\otimes n}=M^{\otimes n}\Lambda^{\otimes n}(N^{\otimes n})^T
\end{equation}
From the uniqueness of the SVD, we know that $M_n=M^{\otimes n}$, $\Lambda_n=\Lambda^{\otimes n}$
and $N_n=N^{\otimes n}$. Then, the ordered singular values of $\tilde{P}_{U^nV^n}$ are
\begin{equation}
\{1, \lambda_2(\tilde{P}_{UV}), \dots, \lambda_2(\tilde{P}_{UV}),\dots\}\nonumber
\end{equation}
where the second through the $n+1$-st singular values are all equal to $\lambda_2(\tilde{P}_{UV})$.

%

From Theorem \ref{sigpro}, we know that if $X_1\rightarrow U^n\rightarrow V^n\rightarrow X_2$ with $n\rightarrow\infty$, then, for $i=2,\dots, \min(|\mathcal{X}_1|,|\mathcal{X}_2|)$,
\begin{equation}
\lambda_i(\tilde{P}_{X_1X_2})\le\lambda_2(\tilde{P}_{X_1U^n})\lambda_i(\tilde{P}_{U^nV^n})\lambda_2(\tilde{P}_{V^nX_2})
\end{equation}
We showed above that $\lambda_i(\tilde{P}_{U^nV^n})\le\lambda_2(\tilde{P}_{UV})$ for $i\ge 2$, and $\lambda_i(\tilde{P}_{U^nV^n})=\lambda_2(\tilde{P}_{UV})$ for $i=2,\dots,n+1$. Therefore,   for $i=2,\dots, \min(|\mathcal{X}_1|,|\mathcal{X}_2|)$, we have
\begin{equation}\label{ine}
\lambda_i(\tilde{P}_{X_1X_2})\le\lambda_2(\tilde{P}_{X_1U^n})\lambda_2(\tilde{P}_{UV})\lambda_2(\tilde{P}_{V^nX_2})
\end{equation}
From Theorem \ref{iff}, we know that $\lambda_2(\tilde{P}_{X_1U^n})\le 1$ and $\lambda_2(\tilde{P}_{V^nX_2})\le 1$.
Next, in  Theorem \ref{app},  we determine that the least upper bound for
$\lambda_2(\tilde{P}_{X_1U^n})$ and $\lambda_2(\tilde{P}_{V^nX_2})$ is also $1$.
\begin{Theo}\label{app}
Let $F(n, P_{X_1})$ be the set of all  joint distributions for $X_1$ and $U^n$
with a given marginal distribution for $X_1$, $P_{X_1}$. Then,
\begin{equation}
\sup_{F(n, P_{X_1}),\; n=1,2,\dots}\lambda_2(\tilde{P}_{X_1U^n})=1\label{mapping}
\end{equation}
\end{Theo}
The proof of Theorem  \ref{app} is given in Appendix \ref{proofs1}.

Based on the above discussion, we have the following theorem.
\begin{Theo} \label{necc}
If $X_1\rightarrow U^n\rightarrow V^n\rightarrow X_2$,
then, for $i=2,\dots, \min(|\mathcal{X}_1|, |\mathcal{X}_2|)$,
\begin{equation}
\lambda_i(\tilde{P}_{X_1X_2})\le\lambda_2(\tilde{P}_{UV})
\end{equation}
\end{Theo}

Theorem \ref{necc} provides a single-letter necessary condition for the $n$-letter Markov chain $X_1\rightarrow U^n\rightarrow V^n\rightarrow X_2$ on the joint probability distribution $p(x_1,x_2)$. This theorem also answers the questions we posed in Section \ref{introPF}. Our first question was whether $(X_1, X_2)$ can be arbitrarily correlated, when $n$ goes to infinity.
Theorem \ref{necc} shows that $(X_1, X_2)$ cannot be arbitrarily correlated, as the correlation measures between $(X_1, X_2)$, $\lambda_i(\tilde{P}_{X_1X_2})$, are upper bounded by, $\lambda_2(\tilde{P}_{UV})$, the second correlation measure of the single-letter sources $(U,V)$. 
Our second question was how much extra correlation $(X_1, X_2)$ can gain when $n$ goes from $1$ to $\infty$.
Although we have no exact answer for this question, the following observation may provide some insights into this problem.
From Theorem \ref{sigpro}, we know that, if $X_1\rightarrow U\rightarrow V\rightarrow X_2$,
\begin{equation}
\lambda_i(\tilde{P}_{X_1X_2})\le\lambda_i(\tilde{P}_{UV}) \qquad i=2,\dots, \min(|\mathcal{X}_1|, |\mathcal{X}_2|)
\end{equation}
Theorem \ref{necc} shows, on the other hand, that, if $X_1\rightarrow U^n\rightarrow V^n\rightarrow X_2$,
\begin{equation}
\lambda_i(\tilde{P}_{X_1X_2})\le\lambda_2(\tilde{P}_{UV}) \qquad i=2,\dots, \min(|\mathcal{X}_1|, |\mathcal{X}_2|)
\end{equation}
Therefore, we note that $n$ going from $1$ to $\infty$ increases the upper bounds\footnote{In general, these upper bounds are not tight.} for the correlation measures $\lambda_i(\tilde{P}_{X_1X_2})$ from $\lambda_i(\tilde{P}_{UV})$ to $\lambda_2(\tilde{P}_{UV})$ for $i=3,\dots, \min(|\mathcal{X}_1|, |\mathcal{X}_2|)$.

As we mentioned in Section \ref{introPF}, the data processing inequality in its usual form  \cite[p. 32]{Cover:1991} is not helpful in this problem, while our new data processing inequality, i.e.,~Theorem \ref{pro}, provides a single-letter necessary condition for this $n$-letter Markov chain. The main reason for this difference is that while the mutual information, $I(U^n; V^n)$,  the correlation measure in the original data processing inequality, increases linearly with $n$, $\lambda_i(\tilde{P}_{U^nV^n})$, the correlation measure in our new data processing inequality, is bounded as $n$ increases, and therefore, makes the problem more tractable.

Theorem \ref{necc} is valid for all discrete random variables. To illustrate the utility and also the limitations of Theorem \ref{necc}, we will study a binary example in detail in Appendix \ref{biexample}. In this example,  $(U,V)$ and $(X_1,X_2)$ are  binary random variables. For this specific binary example, we will apply Theorem \ref{necc} to obtain a necessary condition for the $n$-letter Markov chain. Moreover, the special structure of this binary example will enable us to provide a sharper necessary condition than the one given in Theorem \ref{necc}. We will compare these two necessary conditions and a sufficient condition for this binary example.

\subsection{Conditional Distributions} \label{condD}
Theorem \ref{necc} in Section \ref{iid} provides a necessary condition for  joint probability distributions $p(x_1,x_2)$, which satisfy the Markov chain $X_1\longrightarrow U^n
\longrightarrow V^n \longrightarrow X_2$. 
In certain specific problems, e.g.,~multi-terminal rate-distortion problem and multiple access channel with correlated sources, in addition to $p(x_1,x_2)$, the distributions of $(X_1, X_2)$ conditioned on parts of the $n$-letter sources may be needed, e.g.,~$p(x_1,x_2|u_1,v_1)$, $p(x_1,x_2|u_1,u_2, v_1,v_2)$, etc.\footnote{The reader may wish to consult Sections \ref{MTRD} and \ref{MAC} for further motivations to consider conditional probability distributions.} In this section, we will develop a  result similar to that in Theorem \ref{necc} for conditional distributions.
 
For a pair of i.i.d. sequences $(U^n, V^n)$ of length $n$, we define $\underline{U}$ as an arbitrary subset of $\{U_1,\dots,U_n\}$,
i.e.,
\begin{equation}
\underline{U}\triangleq\{U_{i_1},\dots,U_{i_l}\}\subset \{U_1,\dots, U_n\}
\end{equation} 
and similarly, 
\begin{equation}
\underline{V}\triangleq\{V_{j_1},\dots,V_{j_k}\}\subset \{V_1,\dots,V_n\}\end{equation}
In the following theorem, we propose an upper bound for $\lambda_{i}(\tilde{P}_{X_1X_2|\underline{u}\underline{v}})$, when  $X_1\longrightarrow U^n
\longrightarrow V^n \longrightarrow X_2$ is satisfied.
\begin{Theo}\label{nec}
Let $(U^n, V^n)$ be a pair of i.i.d. sequences of length $n$, and let
the random variables $X_1, X_2$ satisfy $X_1\longrightarrow U^n
\longrightarrow V^n \longrightarrow X_2$. 
Then, for $i=2,\dots,
\min(|\mathcal{X}_1|,|\mathcal{X}_2|)$,
\begin{align}
\lambda_{i}(\tilde{P}_{X_1X_2|\underline{u}\underline{v}})&\le\lambda_2(\tilde{P}_{UV})\label{condi1}
\end{align}
where $\underline{U}\subset \{U_1,\dots, U_n\}$ and $\underline{V}\subset \{V_1,\dots, V_n\}$.
\end{Theo}
\begin{proof}
We consider a special case of $(\underline{U}, \underline{V})$ as follows.
We define $\underline{U}\triangleq\{U_1,\dots, U_l\}$ and
 $\underline{V}\triangleq\{V_1,\dots,V_m, V_{l+1}, \dots, V_{l+k-m}\}$.
 We also define the complements of $\underline{U}$ and $\underline{V}$ as:
$\underline{U}^c\triangleq \{U_{1},\dots,U_n\}\backslash \underline{U}$ and 
$\underline{V}^c\triangleq \{V_{1},\dots,V_n\}\backslash \underline{V}$.
If $\underline{U}$ and $\underline{V}$ take other forms, we can transform them to the form 
we defined above by permutations.
We know that
\begin{align}
p(x_1,x_2, \underline{u}^c,\underline{v}^c|\underline{u},\underline{v})&=p(x_1|\underline{u}^c,\underline{u}, \underline{v})p(\underline{u}^c,\underline{v}^c|\underline{u},\underline{v})p(x_2|\underline{v}^c,\underline{v},\underline{u})
\end{align}
In other words, given $\underline{U}=\underline{u}$ and $\underline{V}=\underline{v}$, $(X_1, \underline{U}^c, \underline{V}^c, X_2)$ form a Markov chain.
Thus,  from (\ref{prod}),
\begin{equation}\label{long}
\tilde{P}_{X_1X_2|\underline{u}\underline{v}}=
\tilde{P}_{X_1\underline{U}^c|\underline{u}\underline{v}}\tilde{P}_{\underline{U}^c\underline{V}^c|\underline{u}\underline{v}}
\tilde{P}_{\underline{V}^cX_2|\underline{u}\underline{v}}
\end{equation}
Furthermore, 
\begin{align}
\tilde{P}_{\underline{U}^c\underline{V}^c|\underline{u}\underline{v}}=&\tilde{p}_{V_{m+1}^l|u_{m+1}^l}^T\otimes
\tilde{p}_{U_{l+1}^{l+k-m}|v_{l+1}^{l+k-m}}
\otimes\tilde{P}_{U_{l+k-m+1}^nV_{l+k-m+1}^n}
\end{align}
As mentioned earlier, a vector marginal distribution can be viewed as a 
joint distribution matrix with a degenerate random variable
whose alphabet size is equal to $1$. Since the rank of a vector is 
$1$, from  Theorem \ref{iff}, the sole singular value of $\tilde{p}_{V_{m+1}^l|u_{m+1}^l}$ (and of
$\tilde{p}_{U_{l+1}^{l+k-m}|v_{l+1}^{l+k-m}}$) is equal to $1$.
Then, 
\begin{equation}\label{comb}
\lambda_{i}(\tilde{P}_{\underline{U}^c\underline{V}^c|\underline{u}\underline{v}})=\lambda_{i}(\tilde{P}_{U_{l+k-m+1}^nV_{l+k-m+1}^n})
\end{equation}
Combining (\ref{newd}), (\ref{long}), and (\ref{comb}), we obtain
\begin{equation}
\lambda_{i}(\tilde{P}_{X_1X_2|\underline{u}\underline{v}})\le \lambda_{2}(\tilde{P}_{UV})
\end{equation}
which completes the proof. 
\end{proof} 

\subsection{General Result}\label{GeneR}
In Sections \ref{iid} and \ref{condD}, we proposed necessary conditions for the $n$-letter Markov chain $X_1\longrightarrow U^n
\longrightarrow V^n \longrightarrow X_2$ on $p(x_1,x_2)$ and $p(x_1,x_2|\underline{u}\underline{v})$, respectively. With these tools, we will develop a general result in this section.   We define the set $\mathcal{S}_{X_1X_2|\mathbf{UV}}$ as follows
\begin{equation}
\mathcal{S}_{X_1X_2|\mathbf{UV}}\triangleq \{p(x_1,x_2|\mathbf{u},\mathbf{v}): X_1\longrightarrow U^n \longrightarrow V^n \longrightarrow X_2, n\rightarrow\infty\}
\end{equation}
where $\mathbf{U}\subset\{U_1,\dots,U_n\}$ and  $\mathbf{V}\subset\{V_1,\dots,V^n\}$. 
We may invoke Theorem \ref{nec} with $(\underline{U},\underline{V})=(\mathbf{U},\mathbf{V})$ and obtain
\begin{align}
\mathcal{S}_{\mathbf{UV}}&\triangleq\{p(x_1,x_2|\mathbf{u},\mathbf{v}):\lambda_{i}(\tilde{P}_{X_1X_2|\mathbf{uv}})
\le\lambda_{2}(\tilde{P}_{UV}), i=1,\dots,\min(|\mathcal{X}_1|,|\mathcal{X}_2|)\}\nonumber\\&\supseteq \mathcal{S}_{X_1X_2|\mathbf{UV}}
\end{align}
In the following, we use Theorem \ref{nec} with different choices of set arguments to find a set that is smaller than $\mathcal{S}_{\mathbf{UV}}$, but still contains $\mathcal{S}_{X_1X_2|\mathbf{UV}}$. 

We note that for a given source distribution $p(u,v)$, we can obtain $p(x_1,x_2|\mathbf{u}',\mathbf{v}')$ (or equivalently $\tilde{P}_{X_1X_2|\mathbf{u'v'}}$) for any $\mathbf{U}'\subseteq\mathbf{U}$ and $\mathbf{V}'\subseteq\mathbf{V}$, from the conditional distribution $p(x_1,x_2|\mathbf{u},\mathbf{v})$. 
Thus, if we define
\begin{equation}
\mathcal{S}_{\mathbf{U'V'}}\triangleq \{p(x_1, x_2|\mathbf{u}, \mathbf{v}): \lambda_{i}(\tilde{P}_{X_1X_2|\mathbf{u'v'}})
\le\lambda_{2}(\tilde{P}_{UV}), i=1,\dots,\min(|\mathcal{X}_1|,|\mathcal{X}_2|)\}
\end{equation}
then, by invoking Theorem \ref{nec} with $(\underline{U},\underline{V})=(\mathbf{U}',\mathbf{V}')$, we have 
\begin{equation}
\mathcal{S}_{X_1X_2|\mathbf{UV}}\subseteq\mathcal{S}_{\mathbf{U'V'}}
\end{equation}
Consequently, if we define
\begin{equation}
\mathcal{S'}_{X_1X_2|\mathbf{UV}}\triangleq \bigcap_{\mathbf{U}'\subseteq \mathbf{U}, \mathbf{V}'\subseteq \mathbf{V}}
 \mathcal{S}_{\mathbf{U'V'}}
\end{equation}
then, we have
\begin{equation}
\mathcal{S}_{X_1X_2|\mathbf{UV}}\subseteq\mathcal{S}_{X_1X_2|\mathbf{UV}}'\subseteq\mathcal{S}_{\mathbf{UV}}
\end{equation}
That is, when we need a necessary condition on $p(x_1,x_2|\mathbf{u},\mathbf{v})$,
even though $\mathcal{S}_{\mathbf{UV}}$ provides such a necessary condition, we can obtain
a smaller probability set and therefore a stricter necessary condition
by combining the necessary conditions for all $p(x_1,x_2|\mathbf{u}',\mathbf{v}')$ where
the sets $\mathbf{U}'$ and $\mathbf{V}'$ are included in the sets $\mathbf{U}$ and $\mathbf{V}$, respectively.

\section{Example I: Multi-terminal Rate-distortion Region}\label{MTRD}
Ever since the milestone paper of Wyner and Ziv \cite{Wyner:1976} on the rate-distortion function of a single source with  side information  at the decoder, there has been a significant amount of efforts directed towards solving a generalization of this problem, the so called multi-terminal rate-distortion problem.   Among all the attempts on this difficult problem,  the notable works by Tung \cite{Tung:1978} and Housewright \cite{Housewright:1977} (see also \cite{Berger:1978}) provide the inner and outer bounds for the rate-distortion region. A more recent progress on this problem is by Wagner and Anantharam in \cite{Wagner:2006}, where a tighter outer bound is given. A very promising and very recent result can be found in \cite{Servetto:2006}.

The multi-terminal rate-distortion problem can be formulated as follows. Consider a pair of discrete memoryless sources $(U, V)$, with joint distribution $p(u,v)$ defined on the finite alphabet $\mathcal{U}\times\mathcal{V}$. 
The reconstruction of the sources are built on another finite alphabet $\hat{\mathcal{U}}\times \hat{\mathcal{V}}$.
The distortion measures are defined as $d_1: \mathcal{U}\times\hat{\mathcal{U}}\longmapsto \mathbb{R}^+\cup\{0\}$ and $d_2: \mathcal{V}\times\hat{\mathcal{V}}\longmapsto \mathbb{R}^+\cup\{0\}$. 
Assume that  two distributed encoders are functions $f_1:\mathcal{U}^n\longmapsto \{1,2,\dots,M_1\}$ and 
$f_2:\mathcal{V}^n\longmapsto \{1,2,\dots,M_2\}$ and a joint decoder is the function $g:\{1,2,\dots,M_1\}
\times \{1,2,\dots,M_2\}\longmapsto\hat{\mathcal{U}}^n\hat{\times\mathcal{V}^n}$, 
where $n$ is a positive integer.
A pair of distortion levels $\mathbf{D}\triangleq(D_1,D_2)$ is said to be $\mathbf{R}$-attainable, for some rate pair
$\mathbf{R}\triangleq(R_1, R_2)$,
if for all $\epsilon>0$ and $\delta>0$, there exist, some positive integer $n$ and a set of distributed encoders and  joint decoder $(f_1, f_2, g)$ with rates $(\frac{1}{n}\log_2 M_1, \frac{1}{n}\log_2 M_2)=(R_1+\delta, R_2+\delta)$,
such that the distortion between the sources $(U^n, V^n)$ and the decoder output $(\hat{U}^n, \hat{V}^n)$
satisfies\footnote{By $(A, B)<(C,D)$, we mean  both $A<B$ and $C<D$, and $(A, B)\le(C,D)$ is defined in the similar manner.}
 $\big(Ed_1(U^n, \hat{V}^n), Ed_2(V^n, \hat{V}^n)\big)<(D_1+\epsilon, D_2+\epsilon)$
where $d_1(U^n, \hat{U}^n)\triangleq\frac{1}{n}\sum_{i=1}^nd_1(U_i,\hat{U}_i)$ and 
$d_2(V^n, \hat{V}^n)\triangleq\frac{1}{n}\sum_{i=1}^nd_2(V_i,\hat{V}_i)$.
The problem here is to determine,  for a fixed $\mathbf{D}$, the set $\mathcal{R}(\mathbf{D})$ of all
rate pairs
$\mathbf{R}$, for which $\mathbf{D}$ is $\mathbf{R}$-attainable.
\subsection{Existing Results}
We restate the outer bound provided in \cite{Tung:1978} and \cite{Housewright:1977}  in the following theorem.
\begin{Theo}\cite{Tung:1978,Housewright:1977}\label{btb}
$\mathcal{R}(\mathbf{D})\subseteq\mathcal{R}_{{out,1}}(\mathbf{D})$, where $\mathcal{R}_{{out,1}}(\mathbf{D})$
is the set of all $\mathbf{R}$ such that there exists a pair  of discrete random variables $(X_1,X_2)$,  for which 
the following three conditions are satisfied:
\begin{enumerate}
\item The joint distribution satisfies
\begin{align}
X_1\rightarrow U&\rightarrow V\\
U&\rightarrow V\rightarrow X_2
\end{align}
\item  The rate pair satisfies
\begin{align}
R_1&\ge I(U,V; X_1|X_2)\\
R_2&\ge I(U,V; X_2|X_1)\\
R_1+R_2&\ge I(U,V; X_1,X_2)
\end{align}
\item There exists $\big(\hat{U}(X_1,X_2), \hat{V}(X_1,X_2)\big)$ such that 
$\big(Ed_1(U, \hat{U}), Ed_2(V, \hat{V}\big))\le \mathbf{D}$.
\end{enumerate}
\end{Theo}
An  inner bound is also given in \cite{Tung:1978} and \cite{Housewright:1977} as follows.
\begin{Theo}\cite{Tung:1978,Housewright:1977}
$\mathcal{R}(\mathbf{D})\supseteq\mathcal{R}_{{in}}(\mathbf{D})$, where $\mathcal{R}_{{in}}(\mathbf{D})$
is the set of all $\mathbf{R}$ such that there exists a pair  of discrete random variables $(X_1,X_2)$,  for which 
the following three conditions are satisfied:
\begin{enumerate}
\item The joint distribution satisfies
\begin{align}
X_1\rightarrow U&\rightarrow V\rightarrow X_2
\end{align}
\item  The rate pair satisfies
\begin{align}
R_1&\ge I(U,V; X_1|X_2)\\
R_2&\ge I(U,V; X_2|X_1)\\
R_1+R_2&\ge I(U,V; X_1,X_2)
\end{align}
\item There exists $\big(\hat{U}(X_1,X_2), \hat{V}(X_1,X_2)\big)$ such that 
$\big(Ed_1(U, \hat{U}), Ed_2(V, \hat{V})\big)\le \mathbf{D}$.
\end{enumerate}
\end{Theo}

We note that the inner and outer bounds agree on both the second condition, i.e.,~the rate constraints in terms of some mutual information expressions, and the third condition, i.e.,~ the reconstruction functions.  However, the first condition in these two bounds constraining the underlying probability distributions $p (x_1, x_2|u, v)$ are different. It is easy to see that the Markov chain condition in the inner bound, i.e.,~$X_1\rightarrow U\rightarrow V\rightarrow X_2$,
implies the Markov chain conditions in the outer bound, i.e.,~$X_1\rightarrow U\rightarrow V$ and $U\rightarrow V\rightarrow X_2$.  Hence, if we define
\begin{align}
\mathcal{S}_{out,1}&\triangleq\{p(x_1,x_2|u,v): X_1\rightarrow U\rightarrow V\text{ and }U\rightarrow V\rightarrow X_2\}\\
 \mathcal{S}_{in}&\triangleq\{p(x_1,x_2|u,v): X_1\rightarrow U\rightarrow V\rightarrow X_2\}
\end{align}
then, 
\begin{equation}
\mathcal{S}_{in}\subseteq\mathcal{S}_{out,1}
\end{equation}
Using the time-sharing argument, a convexification of the inner bound $\mathcal{R}_{in}(\mathbf{D})$ yields another inner bound  $\mathcal{R}_{in}'(\mathbf{D})$, which is larger than $\mathcal{R}_{in}(\mathbf{D})$. This new inner bound may be expressed as a function of $\mathcal{S}_{in}$ and $\mathbf{D}$ as follows,
\begin{equation}
\mathcal{R}_{in}(\mathbf{D})
\subseteq\mathcal{R}'_{in}(\mathbf{D})=\mathcal{F}(\mathcal{S}_{in},\mathbf{D}) \subseteq
\mathcal{R}(\mathbf{D})
\end{equation}
where, using a time sharing random variable $Q$, which is 
 known by the  encoders and the decoder,  $\mathcal{F}(\mathcal{S}_{in},\mathbf{D})$ is defined as,
\begin{align}
\mathcal{F}(\mathcal{S}_{in},\mathbf{D})\triangleq& \bigcup_{\mathbf{p}\in\mathcal{P}(\mathcal{S}_{in},\mathbf{D})}\mathcal{C}(\mathbf{p})\\
\mathbf{p}\triangleq&p(x_1,x_2,q|u,v)=p_q(x_1,x_2|u,v)p(q)\\
\mathcal{P}(\mathcal{S}_{in},\mathbf{D})\triangleq&\left\{\mathbf{p}:\begin{array}{l}
p_q(x_1,x_2|u,v)\in\mathcal{S}_{in};\\
 \exists \big(\hat{U}(X_1,X_2, Q), \hat{V}(X_1,X_2, Q)\big), \\
 \text{ s.t. }\big(Ed_1(U, \hat{U}), Ed_2(V, \hat{V})\big)\le \mathbf{D}
\end{array}\right\}\\
\mathcal{C}(\mathbf{p})\triangleq&\left\{(R_1, R_2):\!\!\!\!\begin{array}{rcl}R_1\ge I(U,V;X_1|X_2, Q)\\R_2\ge I(U,V;X_2|X_1, Q)\\R_1+R_2\ge I(U,V;X_1,X_2|Q)\end{array}\right\}
\end{align}
From the definition of the function $\mathcal{F}$, we can see that $\mathcal{F}$ is monotonic with respect to the set argument when the distortion argument is fixed, i.e.,
\begin{equation}\label{monotonic}
\mathcal{F}(A,\mathbf{D})\subseteq\mathcal{F}(B,\mathbf{D}),\qquad\text{ if }A\subseteq B
\end{equation}

 In \cite{Housewright:1977}, it was shown that $\mathcal{R}_{out,1}(\mathbf{D})$ is convex. Thus,  $\mathcal{R}_{out,1}(\mathbf{D})$  can  be represented in terms of function $\mathcal{F}$ as well, i.e.,
 \begin{equation}
\mathcal{R}_{out,1}(\mathbf{D})=\mathcal{F}(\mathcal{S}_{out,1},\mathbf{D}) 
 \end{equation}
 
The result by Wagner and Anatharam \cite{Wagner:2006} can also be expressed by using the function $\mathcal{F}$ as\footnote{This is a simplified version of \cite{Wagner:2006} with the assumption that there is no hidden source behind $(U^n, V^n)$.}
\begin{equation}\label{WAbo}
\mathcal{R}_{out,2}(\mathbf{D})=\mathcal{F}(\mathcal{S}_{out,2},\mathbf{D}) 
\end{equation}
where
\begin{equation}\label{triMCO}
\mathcal{S}_{{out,2}}\triangleq \{p(x_1,x_2|u,v):\exists w,  p(x_1,x_2,w|u,v)=p(w)p(x_1|w,u)p(x_2|w,v)\}
\end{equation}
The distribution in (\ref{triMCO}) may be represented by the following Markov chain like notation
\begin{equation}\label{triMC}
\begin{array}{llcrr}
X_1&\rightarrow U&\rightarrow& V\rightarrow &X_2\\
&\searrow&&\nearrow&\\
&&W&&
\end{array}
\end{equation}
We note that 
\begin{equation}
\mathcal{S}_{in}
\subseteq
\mathcal{S}_{out,2}
\subseteq
\mathcal{S}_{out,1}
\end{equation} 
Therefore, we conclude that the gap between the inner and the outer bounds comes only from the difference between the feasible sets of the probability distributions $p(x_1,x_2|u,v)$. In the next section, we will provide a tighter outer bound for the rate region in the sense that it can be represented using the same mutual information expressions, however, on a smaller feasible set for $p(x_1,x_2|u,v)$ than $\mathcal{R}_{out,2}(\mathbf{D})$.

\subsection{A New Outer Bound}
We propose a new outer bound for the multi-terminal rate-distortion region as follows.
\begin{Theo}\label{nletter2}
$\mathcal{R}(\mathbf{D})\subseteq\mathcal{R}_{{out,2}}(\mathbf{D})$, where $\mathcal{R}_{{out,2}}(\mathbf{D})$
is the set of all $\mathbf{R}$ such that there exist some positive integer $n$,
 and discrete random variables $Q, X_1, X_2$ for which 
the following three conditions are satisfied:
\begin{enumerate}
\item The joint distribution satisfies
\begin{align}
p(u^n,& v^n, x_1, x_2,  q)
=p(q)p(x_1|u^n, q)p(x_2|v^n, q)\prod_{i=1}^n p(u_i, v_i)\label{dist}
\end{align}
\item The rate pair satisfies
\begin{align}
R_1&\ge I(U_1,V_1; X_1|X_2, Q)\label{r1}\\
R_2&\ge I(U_1,V_1; X_2|X_1, Q)\label{r2}\\
R_1+R_2&\ge I(U_1,V_1; X_1,X_2|Q)\label{r1r2}
\end{align}
where $(U_1, V_1)$ is the first sample of the $n$-sequences $(U^n, V^n)$.
\item There exists $\big(\hat{U}(X_1,X_2, Q), \hat{V}(X_1,X_2, Q)\big)$ such that $\big(Ed_1(U, \hat{U}), Ed_2(V, \hat{V})\big)\le \mathbf{D}$.
\end{enumerate}
or equivalently,
\begin{equation}
\mathcal{R}_{out,3}(\mathbf{D})=\mathcal{F}(\mathcal{S}_{out,3},\mathbf{D}) 
\end{equation}
where
\begin{equation}
\mathcal{S}_{out,3}\triangleq \{p(x_1,x_2|u_1,v_1): X_1\rightarrow U^n\rightarrow V^n\rightarrow X_2\}
\end{equation}
\end{Theo}
\begin{proof}
We consider an arbitrary triple $(f_1, f_2, g)$ 
 of two distributed encoders and one joint decoder with reconstructions $(\hat{U}^n, \hat{V}^n)=g(Y, Z)$, where $Y=f_1(U^n)$ and $Z=f_2(V^n)$,  such that the distortions satisfy $\big(Ed_1(U^n, \hat{V}^n), Ed_2(V^n, \hat{V}^n)\big)<(D_1+\epsilon, D_2+\epsilon)$.  Here, we use $R_1=\frac{1}{n}\log_2(M_1)=\frac{1}{n}\log_2(|Y|)$ and $R_2=\frac{1}{n}\log_2(M_2)=\frac{1}{n}\log_2(|Z|)$.

We define the auxiliary random variables $X_{1i}=(Y, U^{i-1})$ and $X_{2i}=(Z, V^{i-1})$. Then, we have
\begin{align}
\log_2(M_1)&\ge H(Y)\nonumber\\
&=I(U^n,V^n; Y)\nonumber\\
&\overset{1}{\ge} I(U^n,V^n; Y|Z)\nonumber\\
&=\sum_{i=1}^n I(U_i, V_i; Y|Z, U^{i-1}, V^{i-1})\nonumber\\
&= \sum_{i=1}^n I(U_i, V_i; Y,Z| U^{i-1}, V^{i-1}) 
-I(U_i, V_i; Z| U^{i-1}, V^{i-1})\nonumber\\
&\overset{2}{=} \sum_{i=1}^n I(U_i, V_i; Y,Z| U^{i-1}, V^{i-1})
-I(U_i, V_i; Z| V^{i-1})\nonumber\\
&= \sum_{i=1}^n I(U_i, V_i; Y,Z, U^{i-1}|  V^{i-1})
-I(U_i, V_i; U^{i-1}|V^{i-1})-I(U_i, V_i; Z| V^{i-1})\nonumber\\
&\overset{3}{=}\sum_{i=1}^n I(U_i, V_i; Y,Z, U^{i-1}|  V^{i-1})
-I(U_i, V_i; Z| V^{i-1})\nonumber\\
&=\sum_{i=1}^n I(U_i, V_i; Y, U^{i-1}| Z,V^{i-1})\nonumber\\
&=\sum_{i=1}^n I(U_i, V_i;X_{1i}|X_{2i})
\end{align}
where
\begin{enumerate}
\item follows from the fact that 
$Y\rightarrow U^n\rightarrow V^n\rightarrow Z$. 
We observe that the equality holds
 when $Y$ is independent of $Z$;
\item follows from the fact that 
\begin{align}
p(z|u_i,v_i,v^{i-1})&=p(z|u_i,v_i,u^{i-1},v^{i-1})
\end{align}
\item follows from the memoryless property of the sources.
\end{enumerate}
Using a symmetrical argument, we obtain
\begin{equation}
\log_2(M_2)\ge \sum_{i=1}^nI(U_i,V_i; X_{2i}|X_{1i})
\end{equation}
Moreover,
\begin{align}
\log_2(M_1M_2)\ge& H(Y,Z)\nonumber\\=&I(U^n,V^n; Y,Z)\nonumber\\
=&\sum_{i=1}^nH(U_i,V_i)-H(U_i,V_i|Y,Z,U^{i-1},V^{i-1})\nonumber\\=&\sum_{i=1}^n
I(U_i,V_i; X_{1i},X_{2i})
\end{align}

We introduce a time-sharing random variable $Q$, 
which is uniformly distributed on $\{1,\dots, n\}$ and independent of $U^n$ and $V^n$. 
Let the random variables $X_1$ and $X_2$ be such that
\begin{equation}
p(x_{1i},x_{2i}|u_i, v_i, u_i^c, v_i^c)=p(x_1, x_2|u_1, v_1, u_1^c, v_1^c, Q=i)
 \end{equation}
 where $U_i^c\triangleq \{U_1,\dots,U_{i-1}, U_{i+1},\dots, U_n\}$ and $V_i^c$ is defined similarly.
 Then,  
 \begin{align}
 \sum_{i=1}^nI(U_i,V_i; X_{1i}|X_{2i})&=nI(U_1,V_1; X_1|X_2, Q)\\
 \sum_{i=1}^nI(U_i,V_i; X_{2i}|X_{1i})&=nI(U_1,V_1; X_2|X_1, Q)\\
 \sum_{i=1}^nI(U_i,V_i; X_{1i},X_{2i})&=nI(U_1,V_1; X_1,X_2| Q)
 \end{align}
 
The reconstruction pair $(\hat{U},\hat{V})$ is  defined as follows. When $Q=i$, $(\hat{U},\hat{V})\triangleq(\hat{U}_i,\hat{V}_i)$, i.e.,~the $i$-th letter of $(\hat{U}^n, \hat{V}^n)=g(Y, Z)$. $(\hat{U}_i,\hat{V}_i)$ is a function of $(Y,Z)$, and,  therefore, it is  a function of $(X_1, X_2, Q)$. Hence, we have that $(\hat{U},\hat{V})$ is a function of $(X_1, X_2, Q)$, i.e.,~$\big(\hat{U}(X_1, X_2, Q),\hat{V}(X_1, X_2, Q)\big)$.
It is easy to see that  
\begin{align}
\big(Ed_1(U, \hat{U}), Ed_2(V, \hat{V})\big)&=\big(Ed_1(U^n, \hat{V}^n), Ed_2(V^n, \hat{V}^n)\big)
<(D_1+\epsilon, D_2+\epsilon)
\end{align}
which completes the proof. 
\end{proof}

Next, we state and prove that our outer bound given in Theorem \ref{nletter2} is tighter than
$\mathcal{R}_{{out,2}}(\mathbf{D})$ given in (\ref{WAbo}).
\begin{Theo}\label{tighter}
\begin{equation}
\mathcal{R}_{{out,3}}(\mathbf{D})\subseteq
\mathcal{R}_{{out,2}}(\mathbf{D})
\end{equation}
\end{Theo}
\begin{proof}
Here, we provide two proofs. First, 
we prove this theorem by construction. For every $(R_1, R_2)$ point in $\mathcal{R}_{{out,3}}(\mathbf{D})$,
there exist random variables $Q, X_1, X_2$ satisfying (\ref{dist}), $(R_1, R_2)$ pair satisfying (\ref{r1}), (\ref{r2}) and
(\ref{r1r2}), and a reconstruction pair $\big(\hat{U}(X_1,X_2, Q), \hat{V}(X_1,X_2, Q)\big)$ such that $\big(Ed_1(U, \hat{U}), Ed_2(V, \hat{V})\big)\le \mathbf{D}$. According to \cite{Housewright:1977}, let $X_1'=(X_1, Q)	$ and $X_2'=(X_2, Q)$. Then, $p(x_1',x_2'|u_1,v_1)$ belongs to set $\mathcal{S}_{out,2}$. Moreover,
\begin{equation}
R_1\ge I(U,V; X_1|X_2, Q)=I(U,V;X_1'|X_2')
\end{equation}
and similarly,
\begin{equation}
R_2\ge I(U,V; X_2|X_1, Q)=I(U,V;X_2'|X_1')
\end{equation}
and finally,
\begin{align}
R_1+R_2 &\ge I(U,V; X_1,X_2|Q)\nonumber\\
&=H(U, V|Q)-H(U, V|X_1, X_2, Q)\nonumber\\
&\overset{1}{=}H(U, V)-H(U, V|X_1, X_2, Q)\nonumber\\
&= H(U, V)-H(U, V|X_1', X_2')\nonumber\\
&=I(U,V; X_1',X_2')
\end{align}
where
1. follows from the fact that $Q$ is independent of $(U, V)$.
$(\hat{U}, \hat{V})$ is a function of $(X_1, X_2, Q)$, and, therefore,  it is  a function of $(X_1', X_2')=\big((X_1,Q), ( X_2, Q)\big)$.

Hence, for every rate pair $(R_1, R_2)\in \mathcal{R}_{{out,3}}(\mathbf{D})$,
there exist random variables $X_1', X_2'$ such that $p(x_1',x_2'|u_1,v_1)\in \mathcal{S}_{out,2}$,
$(R_1, R_2)$ pair satisfies the mutual information constraints, and the reconstruction satisfies the distortion constraints. In other words, 
$(R_1, R_2)\in \mathcal{R}_{{out,2}}(\mathbf{D})$, proving the theorem.

An alternative proof comes from the comparison of $\mathcal{S}_{out,2}$ and $\mathcal{S}_{out,3}$,   the feasible sets of probability distributions\footnote{In $\mathcal{S}_{out,2}$, the probability distribution is $p(x_1,x_2|u,v)$. Here, we just rename $U=U_1$ and $V=V_1$.} $p(x_1,x_2|u_1,v_1)$.  We note that $X_1\rightarrow U^n\rightarrow V^n\rightarrow X_2$ implies the Markov chain like condition in (\ref{triMC}), which means that 
\begin{equation}
\mathcal{S}_{out,3}\subseteq
\mathcal{S}_{out,2}
\end{equation}
and because of the monotonic property of $\mathcal{F}(\cdot,\mathbf{D})$ in (\ref{monotonic}), we have 
\begin{equation}
\mathcal{F}(\mathcal{S}_{out,3},\mathbf{D})=\mathcal{R}_{{out,3}}(\mathbf{D})\subseteq
\mathcal{R}_{{out,2}}(\mathbf{D})=\mathcal{F}(\mathcal{S}_{out,2},\mathbf{D})
\end{equation}
\end{proof}

\subsection{A New Necessary Condition}
From the proof of Theorem \ref{nletter2}, we note that $(X_{1i}, X_{2i})$ satisfies an $n$-letter Markov chain
constraint $X_{1i}\longrightarrow U^n \longrightarrow V^n \longrightarrow X_{2i}$. From the discussion in Section \ref{GeneR}, we know that if the random variables $X_1$ and $X_2$ satisfy
$X_1\longrightarrow U^n \longrightarrow V^n \longrightarrow X_2$,  
then,
\begin{align}
\lambda_{i}(\tilde{P}_{X_1X_2})&\le\lambda_2(\tilde{P}_{UV})\qquad i=1,\dots,\min(|\mathcal{X}_1|,|\mathcal{X}_2|)\label{cond1}\\
\lambda_{i}(\tilde{P}_{X_1X_2|u_1})&\le\lambda_2(\tilde{P}_{UV})\qquad  i=1,\dots,\min(|\mathcal{X}_1|,|\mathcal{X}_2|)\label{cond2}\\
\lambda_{i}(\tilde{P}_{X_1X_2|v_1})&\le\lambda_2(\tilde{P}_{UV})\qquad  i=1,\dots,\min(|\mathcal{X}_1|,|\mathcal{X}_2|)\label{cond3}\\
\lambda_{i}(\tilde{P}_{X_1X_2|u_1v_1})&\le\lambda_2(\tilde{P}_{UV})\qquad  i=1,\dots,\min(|\mathcal{X}_1|,|\mathcal{X}_2|)\label{cond4}
\end{align}
or equivalently
\begin{equation}
\mathcal{S}_{out,3}\subseteq\mathcal{S}_{out,4}
\end{equation}
where 
\begin{align}
\mathcal{S}_{out,4}\triangleq&\{p(x_1,x_2|u_1,v_1):(\ref{cond1}), (\ref{cond2}), (\ref{cond3}), \text{ and } (\ref{cond4}) \text{ are satisfied}\}
\end{align}
Thus, we have the following theorem
\begin{Theo}
$\mathcal{R}(\mathbf{D})\subseteq\mathcal{R}_{{out,4}}(\mathbf{D})$, where $\mathcal{R}_{{out,4}}(\mathbf{D})$
is the set of all $\mathbf{R}$ such that there exist discrete random variable $Q$ independent of $(U, V)$,  and discrete random variables $X_1, X_2$ for which 
the following three conditions are satisfied:
\begin{enumerate}
\item The joint distribution satisfies,  
\begin{align}
\lambda_{i}(\tilde{P}_{X_1X_2|q})&\le\lambda_2(\tilde{P}_{UV})\qquad  i=1,\dots,\min(|\mathcal{X}_1|,|\mathcal{X}_2|)\\
\lambda_{i}(\tilde{P}_{X_1X_2|uq})&\le\lambda_2(\tilde{P}_{UV})\qquad  i=1,\dots,\min(|\mathcal{X}_1|,|\mathcal{X}_2|)\\
\lambda_{i}(\tilde{P}_{X_1X_2|vq})&\le\lambda_2(\tilde{P}_{UV})\qquad  i=1,\dots,\min(|\mathcal{X}_1|,|\mathcal{X}_2|)\\
\lambda_{i}(\tilde{P}_{X_1X_2|uvq})&\le\lambda_2(\tilde{P}_{UV})\qquad  i=1,\dots,\min(|\mathcal{X}_1|,|\mathcal{X}_2|)
\end{align}
\item The rate pair satisfies
\begin{align}
R_1&\ge I(U,V; X_1|X_2, Q)\\
R_2&\ge I(U,V; X_2|X_1, Q)\\
R_1+R_2&\ge I(U,V; X_1,X_2|Q)
\end{align}
\item There exists $\big(\hat{U}(X_1,X_2, Q), \hat{V}(X_1,X_2, Q)\big)$ such that $\big(Ed_1(U, \hat{U}), Ed_2(V, \hat{V})\big)\le \mathbf{D}$.
\end{enumerate}
Equivalently,
\begin{equation}
\mathcal{R}_{out,4}(\mathbf{D})=\mathcal{F}(\mathcal{S}_{out,4},\mathbf{D}) 
\end{equation}
\end{Theo}
From Section \ref{GeneR}, we have that 
\begin{equation}
\mathcal{S}_{out,3}\subseteq\mathcal{S}_{out,4}
\end{equation}
and therefore
\begin{equation}
\mathcal{R}_{{out,3}}(\mathbf{D})=\mathcal{F}(\mathcal{S}_{out,3},\mathbf{D}) \subseteq\mathcal{R}_{{out,4}}(\mathbf{D})=\mathcal{F}(\mathcal{S}_{out,4},\mathbf{D}) 
\end{equation}
From Theorem \ref{tighter}, we know that
\begin{equation}
\mathcal{S}_{out,3}\subseteq\mathcal{S}_{out,2}
\end{equation}
and
\begin{equation}
\mathcal{R}_{{out,3}}(\mathbf{D})=\mathcal{F}(\mathcal{S}_{out,3},\mathbf{D}) \subseteq\mathcal{R}_{{out,2}}(\mathbf{D})=\mathcal{F}(\mathcal{S}_{out,2},\mathbf{D}) 
\end{equation}

\begin{figure}
\centering
\includegraphics[width=6.5in]{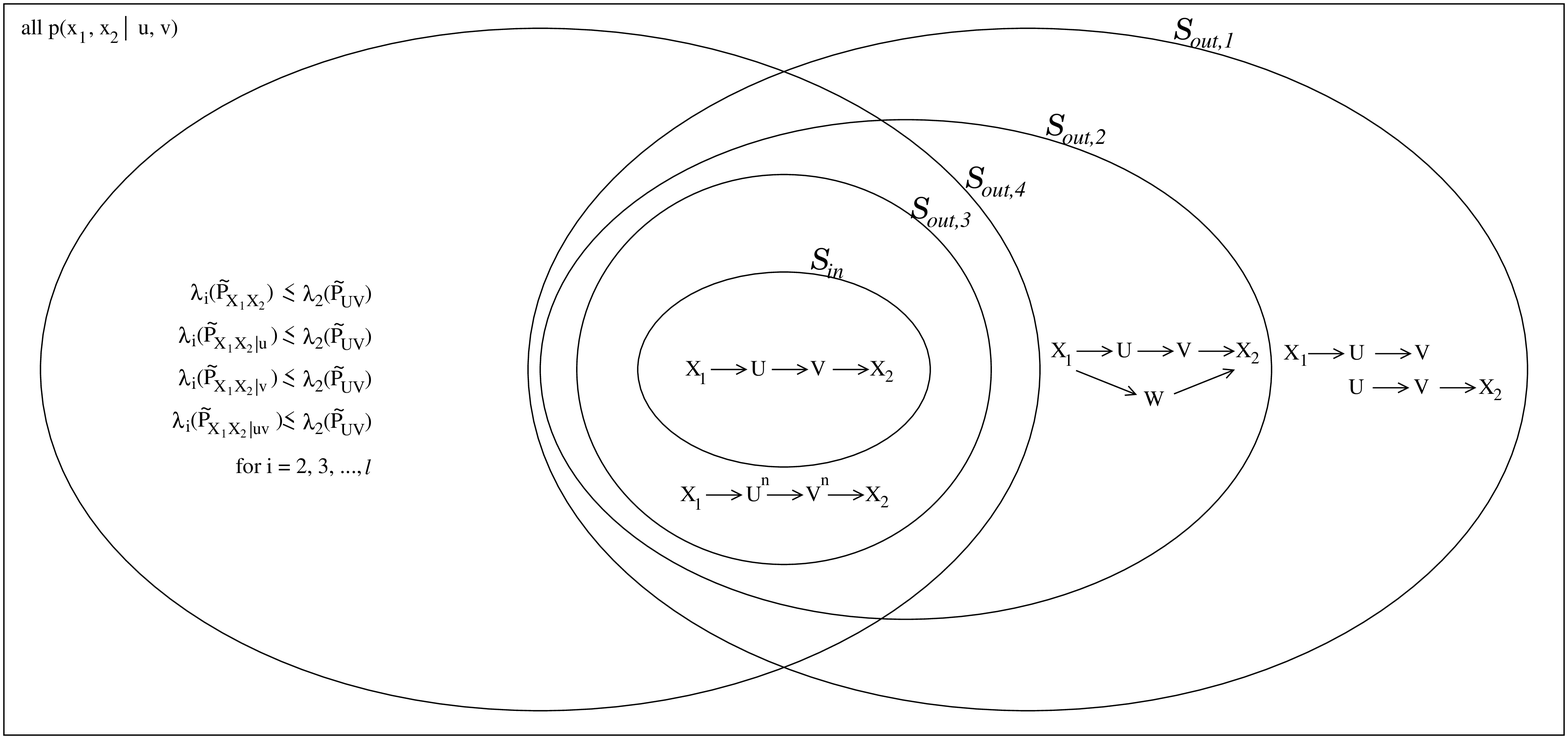}
\caption{Different sets of probability distributions $p(x_1,x_2|u,v)$.}\label{prob_set}
\end{figure}

So far, we have not been able to determine whether
$\mathcal{S}_{out,4}\subseteq\mathcal{S}_{out,2}$ or $\mathcal{S}_{out,2}\subseteq\mathcal{S}_{out,4}$, 
however, we know that there exists some probability distribution $p(x_1,x_2|u_1,v_1)$, which belongs to $\mathcal{S}_{out,2}$, but does not belong to $\mathcal{S}_{out,4}$. For example, assume $\lambda_2(\tilde{P}_{UV})<1$ and some random variable $W$ independent to $(U,V)$. Let $X_1=(f_1(U_1), W)$ and $X_2=(f_2(V_1), W)$. We note that $(X_1, X_2, U_1,V_1)$ satisfies the Markov chain like condition in (\ref{triMC}), i.e.,~ $p(x_1,x_2|u_1,v_1)\in \mathcal{S}_{out,2}$. But, $(X_1, X_2)$ contains common information $W$, which means that $\lambda_2(\tilde{P}_{X_1X_2})=1> \lambda_2(\tilde{P}_{UV})$ \cite{Witsenhausen:1975}, and therefore, $p(x_1,x_2|u_1,v_1)\notin \mathcal{S}_{out,4}$.  Based on this observation,  we note that introducing $\mathcal{S}_{out,4}$ helps us rule out some unachievable probability distributions that may exist in $\mathcal{S}_{out,2}$. 
The relation between different feasible sets of probability distributions $p(x_1,x_2|u_1,v_1)$ is illustrated in  Figure \ref{prob_set}.

Finally, we note that we can obtain a tighter outer bound in terms of the function $\mathcal{F}(\cdot, \mathbf{D})$ by using a set argument which is the intersection of $\mathcal{S}_{out,2}$ and $\mathcal{S}_{out,4}$, i.e.,
\begin{align}
\mathcal{R}_{out, 2\cap 4}(\mathbf{D})&\triangleq\mathcal{F}(\mathcal{S}_{out,2}\cap\mathcal{S}_{out,4}, \mathbf{D})
\end{align}
It is straightforward to see that this outer bound $\mathcal{R}_{out, 2\cap 4}(\mathbf{D})$ is in general tighter than the outer bound $\mathcal{F}(\mathcal{S}_{out,2}, \mathbf{D})$.

\section{Example II: Multiple Access Channel with Correlated Sources}\label{MAC}

The problem of determining the capacity region of the multiple access
channel with correlated sources can be formulated as follows.  Given a
pair of i.i.d. correlated sources $(U,V)$ described by the joint probability
distribution $p(u,v)$, and a discrete, memoryless, multiple access
channel characterized by the transition probability $p(y|x_1, x_2)$,
what are the necessary and sufficient conditions for the reliable
transmission of $n$ 
samples of the sources through the channel, in $n$ channel uses, as
$n\rightarrow\infty$?
\subsection{Existing Results}
The multiple access channel with correlated sources was studied by Cover, El Gamal and Salehi in
\cite{Cover:1980} (a simpler proof was given in \cite{Ahlswede:1983}), where an achievable region expressed by
single-letter entropies and mutual informations was given as follows.
\begin{Theo}\cite{Cover:1980}
A source $(U,V)$ with joint distribution $p(u,v)$ can be sent with arbitrarily small probability
of error over a multiple access channel characterized by $p(y|x_1, x_2)$, if 
there exist probability mass functions $p(s)$, $p(x_1|u,s)$, $p(x_2|v,s)$, such that
\begin{align}
H(U|V)&<I(X_1;Y|X_2, V, S)\\
H(V|U)&<I(X_2;Y|X_1, U, S)\\
H(U,V|W)&<I(X_1, X_2;Y|W,S)\\
H(U, V)&<I(X_1, X_2;Y)
\end{align}
where
\begin{align}
p(s,u,v,&x_1,x_2,y)
=p(s)p(u,v)p(x_1|u,s)p(x_2|v,s)p(y|x_1,x_2)
\end{align}
and
\begin{equation}
w=f(u)=g(v)
\end{equation} 
is the common information in the sense of Witsenhausen, Gacs and Korner (see \cite{Witsenhausen:1975}).
\end{Theo}
The above region can be simplified if there is no common information between $U$ and $V$ as follows \cite{Cover:1980}
\begin{align}
H(U|V)&<I(X_1;Y|X_2, V)\label{ach1}\\
H(V|U)&<I(X_2;Y|X_1, U)\label{ach2}\\
H(U, V)&<I(X_1, X_2;Y)\label{ach3}
\end{align}
where 
\begin{equation}\label{smc}
p(u,v,x_1,x_2,y)
=p(u,v)p(x_1|u)p(x_2|v)p(y|x_1,x_2)
\end{equation}
 This achievable region
was shown to be suboptimal by Dueck \cite{Dueck:1981}.

Cover, El
Gamal and Salehi \cite{Cover:1980} also provided a capacity result
with both achievability and converse in
the form of some incomputable $n$-letter mutual informations. Their result is restated
in the following theorem.

\begin{Theo}\cite{Cover:1980}
The correlated sources $(U, V)$ can be communicated reliably over the
discrete memoryless multiple access channel $p(y|x_1, x_2)$ if and only if
\begin{equation}
[H(U|V), H(V|U), H(U, V)]\in\bigcup_{n=1}^{\infty}\mathcal{C}_n
\end{equation}
where
\begin{equation}
\mathcal{C}_n=\left\{[R_1, R_2, R_3]:
\begin{array}{lll}
\!\!R_1&\!\!\!\!<&\!\!\!\!\frac{1}{n}I(X_1^n;Y^n|X_2^n, V^n)\\
\!\!R_2&\!\!\!\!<&\!\!\!\!\frac{1}{n}I(X_2^n;Y^n|X_1^n, U^n)\\
\!\!R_3&\!\!\!\!<&\!\!\!\!\frac{1}{n}I(X_1^n, X_2^n;Y^n)\\
\end{array}\right\}
\end{equation}
for some 
\begin{align}
p(u^n,&v^n, x_1^n, x_2^n, y^n)=
p(x_1^n|u^n)p(x_2^n|v^n)\prod_{i=1}^n p(u_i, v_i)\prod_{i=1}^n p(y_i|x_{1i}, 
x_{2i})
\label{mc}
\end{align}
i.e., for some $X_1^n$ and $X_2^n$ that satisfy the Markov chain 
$X_1^n \rightarrow U^n \rightarrow V^n \rightarrow X_2^n$.
\end{Theo}

Some recent results on the transmission of correlated sources over multiple access channels can be found in \cite{Pradhan:2006, Lapidoth:2006}.
\subsection{A New Outer Bound}
We propose a new outer bound for the multiple access channel with correlated sources as follows.
\begin{Theo}\label{nletter}
If a pair of i.i.d. sources $(U,V)$ with joint distribution $p(u,v)$
can be transmitted reliably through a discrete, memoryless, multiple
access channel characterized by $p(y|x_1,x_2)$, then
\begin{align}
H(U|V)&\le I(X_{1}; Y|X_{2}, \mathbf{U}, Q)\\
H(V|U)&\le I(X_{2}; Y|X_{1}, \mathbf{V}, Q)\\
H(U,V)&\le I(X_1,X_2;Y|Q)
\end{align}
where random variables $X_1$, $X_2$ and
$Q$ are such that
\begin{align}
p(x_1,& x_2, y, u^n, v^n, q)
=p(q)p(x_1|u^n, q)p(x_2|v^n, q)p(y|x_1, x_2)\prod_{i=1}^n p(u_i, v_i)
\end{align}
where $(U^n, V^n)$ are $n$ samples of the i.i.d. sources with $n\rightarrow\infty$, $\mathbf{U}\subset\{U_1,\dots,U_n\}$ and $\mathbf{V}\subset\{V_1,\dots,V_n\}$ and both $\mathbf{U}$ and $\mathbf{V}$ contain finite number of elements.
\end{Theo}

\begin{proof}
Consider a given block code of length $n$ with the encoders $f_1:
\mathcal{U}^n\longmapsto \mathcal{X}_1^n$ and $f_2:
\mathcal{V}^n\longmapsto \mathcal{X}_2^n$ and decoder $g:
\mathcal{Y}^n\longmapsto \mathcal{U}^n\times\mathcal{V}^n$. From
Fano's inequality \cite[p. 39]{Cover:1991}, we have
\begin{equation}
H(U^n,V^n|Y^n)\le n\log_2|\mathcal{U}\times\mathcal{V}|P_e+1\triangleq 
n\epsilon_n\label{fano}
\end{equation}



Let $G_i$ be a permutation on the set $\{1,\dots,n\}$ (similarly on the set $\{U_1,\dots,U_n\}$, and $\{V_1,\dots,V_n\}$). We define\footnote{For example, if we let $\mathbf{U}=\{U_1, U_2\}$ and $\mathbf{V}=\{V_1,V_2\}$ and $G_1(1)=3$ and $G_1(2)=5$, then, $\mathbf{U}_1=\{U_3, U_5\}$ and $\mathbf{V}_1=\{V_3, V_5\}$.}
\begin{align}
\mathbf{U}_i&\triangleq \{G_i(U_k):U_k\in\mathbf{U}\}\\
\mathbf{V}_i&\triangleq\{G_i(V_k):V_k\in\mathbf{V}\}
\end{align}
This definition provides that $p(\mathbf{u}_i,\mathbf{v}_i)$, the joint probabilities of $\mathbf{U}_i$ and $\mathbf{V}_i$, are identical for $i=1,\dots,n$.

For a code, for which $P_e\rightarrow 0$,  as $n\rightarrow \infty$, we have
$\epsilon_n\rightarrow 0$. 
Then,
\begin{align}
nH(U|V)&=H(U^n|V^n)\nonumber\\
&=I(U^n;Y^n|V^n)+H(U^n|Y^n, V^n)\nonumber\\
&\le I(U^n;Y^n|V^n)+H(U^n, V^n|Y^n)\nonumber\\
&\overset{1}{\le} I(U^n;Y^n|V^n)+ n\epsilon_n \nonumber\\
&=H(Y^n|V^n)-H(Y^n|U^n, V^n)+ n\epsilon_n \nonumber\\
&\overset{2}{=} H(Y^n|X_2^n, V^n)-H(Y^n|X_1^n, X_2^n, U^n, V^n)+ n\epsilon_n\nonumber\\
&\overset{3}{=} H(Y^n|X_2^n, V^n)-H(Y^n|X_1^n, X_2^n)+ n\epsilon_n\nonumber\\
&\overset{4}{=} \sum_{i=1}^n \Big{[}H(Y_i|X_2^n, V^n, Y^{i-1})-H(Y_i|X_{1i}, X_{2i})\Big]+ n\epsilon_n\nonumber\\
&\overset{5}{\le} \sum_{i=1}^n \Big[H(Y_i|X_{2i}, \mathbf{V}_{i})-H(Y_i|X_{1i}, X_{2i})\Big]+ n\epsilon_n\nonumber\\
&\overset{6}{=} \sum_{i=1}^n \Big[H(Y_i|X_{2i}, \mathbf{V}_{i})-H(Y_i|X_{1i}, X_{2i}, \mathbf{V}_{i})\Big]+ n\epsilon_n\nonumber\\
&= \sum_{i=1}^n I(X_{1i};Y_i|X_{2i}, \mathbf{V}_{i})+ n\epsilon_n\label{ub1}
\end{align}
where
\begin{enumerate}
\item from Fano's inequality in (\ref{fano});
\item from the fact that $X_1^n$ is the deterministic function of $U^n$ and 
$X_2^n$ is the deterministic function of $V^n$;
\item from $p(y^n|x_1^n, x_2^n, u^n, v^n)=p(y^n|x_1^n, x_2^n)$;
\item from the chain rule and the memoryless nature of the channel;
\item from the property that conditioning reduces entropy;
\item from $p(y_i|x_{1i}, x_{2i}, \mathbf{v}_i)=p(y_i|x_{1i}, x_{2i})$.
\end{enumerate}
Using a symmetrical argument, we obtain
\begin{equation}
nH(V|U)\le \sum_{i=1}^n I(X_{2i};Y_i|X_{1i}, \mathbf{U}_{i})+ n\epsilon_n\label{ub2}
\end{equation}
Moreover, 
\begin{align}
nH(U,V)&=H(U^n, V^n)\nonumber\\
&=I(U^n, V^n;Y^n)+H(U^n, V^n|Y^n)\nonumber\\
&\le I(U^n, V^n;Y^n)+ n\epsilon_n \nonumber\\
&\le I(X_1^n, X_2^n; Y^n)+ n\epsilon_n \nonumber\\
&=H(Y^n)-H(Y^n|X_1^n, X_2^n)+ n\epsilon_n \nonumber\\
&=\sum_{i=1}^n \Big[H(Y_i|Y^{i-1})-H(Y_i|X_{1i},X_{2i})\Big]+ n\epsilon_n \nonumber\\
&\le \sum_{i=1}^n \Big[H(Y_i)-H(Y_i|X_{1i},X_{2i})\Big]+ n\epsilon_n \nonumber\\
&= \sum_{i=1}^n I(X_{1i},X_{2i};Y_i)+ n\epsilon_n\label{ub3}
\end{align} 
We introduce a time-sharing random variable $Q$
\cite[p. 397]{Cover:1991} as follows. Let $Q$ be uniformly distributed
on $\{1,\dots, n\}$ and be independent of $U^n$, $V^n$.  
Let the random variables $X_1$ and $X_2$ be such that
\begin{align}
p(x_{1i},x_{2i}|\mathbf{u}_i, \mathbf{v}_i,& \mathbf{u}_i^c, \mathbf{v}_i^c)
=p(x_1, x_2|\mathbf{u}, \mathbf{v}, \mathbf{u}^c, \mathbf{v}^c, Q=i)
 \end{align}
where 
\begin{align}
\mathbf{U}^c&\triangleq \{U_1,\dots,U_n\}\backslash \mathbf{U}\\
\mathbf{V}^c&\triangleq \{V_1,\dots,V_n\}\backslash \mathbf{V}\\
\mathbf{U}^c_i&\triangleq \{G_i(U_k):U_k\in\mathbf{U}^c\}\\
\mathbf{V}^c_i&\triangleq \{G_i(V_k):V_k\in\mathbf{V}^c\}
\end{align}
Then, 
\begin{align}
\sum_{i=1}^n I(X_{1i};Y_i|X_{2i}, \mathbf{V}_{i}) & =nI(X_1; Y|X_2, \mathbf{V}, Q)\label{ts1}\\
\sum_{i=1}^n I(X_{2i};Y_i|X_{1i}, \mathbf{U}_{i}) & =nI(X_2; Y|X_1, \mathbf{U}, Q)\label{ts2}\\
\sum_{i=1}^n I(X_{1i},X_{2i};Y_i) & =nI(X_1,X_2; Y|Q)\label{ts3}
\end{align}
Combining (\ref{ts1}), (\ref{ts2}) and (\ref{ts3}) 
with (\ref{ub1}), (\ref{ub2}) and (\ref{ub3}) completes
the proof.
\end{proof}

\subsection{A New Necessary Condition}\label{nc}
It can be shown that the outer bound in Theorem~\ref{nletter} is
equivalent to the following
\begin{equation}
\mathbf{H}\in\mathcal{R}(\mathcal{S})\triangleq 
\mathrm{co}\Big\{\bigcup_{\mathbf{p}\in \mathcal{S}_{X_1X_2|\mathbf{UV}}}\mathcal{R}(\mathbf{p})
\Big\}
\end{equation}
where
\begin{align}
\mathbf{H}&\triangleq[H(U|V), H(V|U), H(U, V)]\\
\mathbf{p}&\triangleq p(x_1, x_2|\mathbf{u}, \mathbf{v})\label{p}\\
\mathcal{S}_{X_1X_2|\mathbf{UV}}&\triangleq \{\mathbf{p}: X_1\longrightarrow U^n \longrightarrow V^n \longrightarrow X_2, n\rightarrow\infty\}\\
\mathcal{R}(\mathbf{p})&\triangleq\left\{[R_1, R_2, R_3]:
\begin{array}{lll}
\!\!R_1&\!\!\!\!\le& \!\!\!\!I(X_{1}; Y|X_{2}, \mathbf{V})\\
\!\!R_2&\!\!\!\!\le& \!\!\!\!I(X_{2}; Y|X_{1}, \mathbf{U})\\
\!\!R_3&\!\!\!\!\le& \!\!\!\!I(X_1,X_2;Y)
\end{array}\right\}\label{mutu}
\end{align}
and $\mathrm{co}\{\cdot\}$ represents the closure of the convex hull
of the set argument.

From Section \ref{GeneR}, we know that
\begin{equation}
\mathcal{S}_{X_1X_2|\mathbf{UV}}\subseteq
\mathcal{S'}_{X_1X_2|\mathbf{UV}}\triangleq \bigcap_{\mathbf{U}'\subseteq \mathbf{U}, \mathbf{V}'\subseteq \mathbf{V}}
 \mathcal{S}_{\mathbf{U'V'}}
\end{equation}
where 
\begin{equation}
\mathcal{S}_{\mathbf{U'V'}}\triangleq \{p(x_1, x_2|\mathbf{u}, \mathbf{v}): \lambda_{i}(\tilde{P}_{X_1X_2|\mathbf{u'v'}})
\le\lambda_{2}(\tilde{P}_{UV}), i=1,\dots,\min(|\mathcal{X}_1|,|\mathcal{X}_2|)\}
\end{equation}

Then, we obtain a single-letter outer bound for the multiple access channel with correlated sources as follows.
\begin{Theo}\label{mare}
If a pair of i.i.d. sources $(U,V)$ with joint distribution $p(u,v)$
can be transmitted reliably through a discrete, memoryless, multiple
access channel characterized by $p(y|x_1, x_2)$, then
\begin{align}
H(U|V)&\le I(X_{1}; Y|X_{2}, \mathbf{V}, Q)\label{conv1}\\
H(V|U)&\le I(X_{2}; Y|X_{1}, \mathbf{U}, Q)\label{conv2}\\
H(U,V)&\le I(X_1,X_2;Y|Q)\label{conv3}
\end{align}
where $\mathbf{U}\subset\{U_1,\dots,U_n\}$ and $\mathbf{V}\subset\{V_1,\dots,V_n\}$ are two sets containing finite letters of source samples,
random variable $Q$ independent of
$(\mathbf{U},\mathbf{V})$, and for random variables $X_1$, $X_2$, 
$p(x_1, x_2|\mathbf{u}, \mathbf{v}, q)$ such that,  for any 
$\mathbf{U}'\subseteq \mathbf{U}$ and $\mathbf{V}'\subseteq \mathbf{V}$,
\begin{align}
\lambda_{i}(\tilde{P}_{X_1X_2|\mathbf{u'v'}q})&\le\lambda_2(\tilde{P}_{UV}),\qquad i=1,\dots,\min(|\mathcal{X}_1|,|\mathcal{X}_2|)\label{fini}
\end{align}
Equivalently,
\begin{equation}
\mathbf{H}\in\mathcal{R}(\mathcal{S'})\triangleq
\mathrm{co}\Big\{\bigcup_{\mathbf{p}\in \mathcal{S'}_{X_1X_2|\mathbf{UV}}}\mathcal{R}(\mathbf{p})\Big\}
\end{equation}
\end{Theo}

In the rest of this section, we will specialize our results to the case where we choose $\mathbf{U}=\{U_1\}$ and $\mathbf{V}=\{V_1\}$. Here, we have the following definitions\footnote{The notation $\mathcal{S}_{out,3}$, as well as $\mathcal{S}_{out,4}$ and $\mathcal{S}_{in}$ in the sequel, is used in order to be consistent with the notations in Section \ref{MTRD}.}
\begin{equation}
\mathcal{S}_{out,3}\triangleq\mathcal{S}_{X_1X_2|U_1V_1}=\{p(x_1,x_2|u_1,v_1):X_1\longrightarrow U^n \longrightarrow V^n \longrightarrow X_2\}
\end{equation}
and
\begin{equation}
\mathcal{S}_{out,4}=\mathcal{S}_{\emptyset}\cap \mathcal{S}_{U_1}\cap \mathcal{S}_{V_1}\cap \mathcal{S}_{U_1V_1}
\end{equation}
where 
\begin{align}
\mathcal{S}_{\emptyset}&\triangleq \{p(x_1, x_2|u_1, v_1): \lambda_{i}(\tilde{P}_{X_1X_2})\le\lambda_2(\tilde{P}_{UV})\}\label{m11}\\
\mathcal{S}_{U_1}&\triangleq \{p(x_1, x_2|u_1, v_1): \lambda_{i}(\tilde{P}_{X_1X_2|u_1})\le\lambda_2(\tilde{P}_{UV})\}\label{m12}\\
\mathcal{S}_{V_1}&\triangleq \{p(x_1, x_2|u_1, v_1): \lambda_{i}(\tilde{P}_{X_1X_2|v_1})\le\lambda_2(\tilde{P}_{UV})\}\label{m13}\\
\mathcal{S}_{U_1V_1}&\triangleq \{p(x_1, x_2|u_1, v_1): \lambda_{i}(\tilde{P}_{X_1X_2|u_1v_1})\le\lambda_2(\tilde{P}_{UV})\}\label{m14}
\end{align}


We note that when $\mathbf{U}=\{U_1\}$ and $\mathbf{V}=\{V_1\}$, the expressions in (\ref{mutu}) 
agree with those in 
the achievability scheme of Cover, El Gamal and Salehi when there is no common information,
i.e.,~(\ref{ach1}), (\ref{ach2}), and (\ref{ach3}). Thus, the gap between the achievablity scheme of Cover, El Gamal and Salehi,
and the converse in this paper results from the fact that the feasible sets for  the conditional 
 probability distribution $\mathbf{p}=p(x_1,x_2|u,v)$ are different.
In the achievability scheme of Cover, El Gamal and Salehi, $\mathbf{p}$ belongs to 
\begin{equation}
\mathcal{S}_{in}\triangleq \{p(x_1,x_2|u,v): X_1\longrightarrow U \longrightarrow V \longrightarrow X_2\}
\end{equation}
since for the achievability, we need $X_1\longrightarrow U \longrightarrow V \longrightarrow X_2$.
Whereas, in our converse, $\mathbf{p}\in \mathcal{S}_{out,3}\subseteq \mathcal{S}_{out,4}$. 
Since $ X_1\longrightarrow U \longrightarrow V \longrightarrow X_2$ implies $ X_1\longrightarrow U^n \longrightarrow V^n \longrightarrow X_2$ and $ X_1\longrightarrow U^n \longrightarrow V^n \longrightarrow X_2$ implies $\lambda_{i}(\tilde{P}_{X_1X_2})\le\lambda_2(\tilde{P}_{UV})$, $\lambda_{i}(\tilde{P}_{X_1X_2|u_1})\le\lambda_2(\tilde{P}_{UV})$, $\lambda_{i}(\tilde{P}_{X_1X_2|v_1})\le\lambda_2(\tilde{P}_{UV})$, and $\lambda_{i}(\tilde{P}_{X_1X_2|u_1v_1})\le\lambda_2(\tilde{P}_{UV})$, 
we have
\begin{equation}
\mathcal{S}_{in}\subseteq \mathcal{S}_{out,3}\subseteq \mathcal{S}_{out,4}
\end{equation}
Therefore, when $m=1$, even though the mutual information expressions in the
achievability and the converse are the same, their actual values will be different, since they will be evaluated 
using the conditional probability distributions that belong to different feasible sets.

\section{Conclusion} 
In the distributed coding on correlated sources, 
the problem of describing a joint distribution involving an $n$-letter Markov chain arises. 
By means of spectrum analysis, we provided a new data processing inequality
based on a new measure of correlation, which gave us a single-letter necessary condition
for the $n$-letter Markov chain. 
We applied our results to two specific examples involving distributed coding of correlated sources:
the multi-terminal rate-distortion region and the multiple access channel with correlated sources, and
proposed two new outer bounds for these two problems.

\appendix
\vspace{0.5in}
\noindent{\Huge \textbf{Appendices}}
\section{An Illustrative Binary Example}\label{biexample}

In this section, we will study a specific binary example in detail. The aims of this study are, first, to ilustrate the single-letter necessary condition we proposed for the $n$-letter Markov chain  in Section \ref{iid}, second, to develop a sharper necessary condition in this specific case, and finally, to compare different necessary conditions and a sufficient condition in this specific example.  

The binary example under consideration is as follows.  Let $U$, $V$, $X_1$ and $X_2$ be binary random variables,
which take values from $\{0,1\}$. We assume that $(U,V)$ are a pair of binary symmetric sources, i.e.,
\begin{equation}
Pr(U=0)=Pr(U=1)=Pr(V=0)=Pr(V=1)=\frac{1}{2}
\end{equation} 
From (\ref{def}) and (\ref{fun}), we have 
\begin{equation}\label{SVDbi}
\tilde{P}_{UV}=\left[\begin{array}{c}\frac{1}{\sqrt{2}}\\\frac{1}{\sqrt{2}}\end{array}\right]
\left[\begin{array}{cc}\frac{1}{\sqrt{2}}&\frac{1}{\sqrt{2}}\end{array}\right]+
\lambda_2(\tilde{P}_{UV})\bm{\mu}_2(\tilde{P}_{UV})\bm{\nu}_2(\tilde{P}_{UV})^T
\end{equation}
Here we focus on the symmetric case, i.e.,
\begin{equation}
\bm{\mu}_2(\tilde{P}_{UV})=\bm{\nu}_2(\tilde{P}_{UV})=\begin{bmatrix}\frac{1}{\sqrt{2}}\\-\frac{1}{\sqrt{2}}\end{bmatrix}
\end{equation} 
In addition, we assume the following marginal distributions for $X_1$ and $X_2$,
\begin{align}
p_{X_1}&=\left[\begin{array}{c}a^2\\1-a^2\end{array}\right]\label{marab1}\\
p_{X_2}&=\left[\begin{array}{c}b^2\\1-b^2\end{array}\right]\label{marab2}
\end{align}
where $0\le a,b\le 1$.
Then, from (\ref{def}) and (\ref{fun}), we have
\begin{align}
\tilde{P}_{X_1X_2}
=&\left[\begin{array}{c}a\\\sqrt{1-a^2}\end{array}\right]
\left[\begin{array}{cc}b&\sqrt{1-b^2}\end{array}\right]+
\lambda_2(\tilde{P}_{X_1X_2})\bm{\mu}_2(\tilde{P}_{X_1X_2})\bm{\nu}_2(\tilde{P}_{X_1X_2})^T
\end{align}
We note that
\begin{equation}
\bm{\mu}_2(\tilde{P}_{X_1X_2})\bm{\nu}_2(\tilde{P}_{X_1X_2})^T
=\sigma\left[\begin{array}{c}\sqrt{1-a^2}\\-a\end{array}\right]\left[\begin{array}{cc}\sqrt{1-b^2}&-b\end{array}\right]
\end{equation}
where $\sigma\in\{1, -1\}$. For the simplicity of the derivation in the sequel, we let $\lambda=\sigma\lambda_{2}(\tilde{P}_{X_1,X_2})$. Then, we have
\begin{equation}
\tilde{P}_{X_1X_2}
=\left[\begin{array}{c}a\\\sqrt{1-a^2}\end{array}\right]
\left[\begin{array}{cc}b&\sqrt{1-b^2}\end{array}\right]+
\lambda
\left[\begin{array}{c}\sqrt{1-a^2}\\-a\end{array}\right]\left[\begin{array}{cc}\sqrt{1-b^2}&-b\end{array}\right]
\end{equation} 

From Theorem \ref{iff}, we know that the entries of $\tilde{P}_{X_1X_2}$ are non-negative, i.e.,
\begin{equation}
\tilde{P}_{X_1X_2}=\begin{bmatrix}ab+\lambda\sqrt{(1-a^2)(1-b^2)}&a\sqrt{1-b^2}-\lambda b\sqrt{1-a^2}\\b\sqrt{1-a^2}-\lambda a\sqrt{1-b^2}&\sqrt{(1-a^2)(1-b^2)}+\lambda ab\end{bmatrix}\ge 0
\end{equation}
which implies that
\begin{equation}
-\xi_2\le\lambda\le
\xi_1
\end{equation}
where
\begin{align}
\xi_1&\triangleq \frac{\min(a^2,b^2)\min(1-a^2,1-b^2)}{ab\sqrt{(1-a^2)(1-b^2)}}\le 1\\
\xi_2&\triangleq \frac{\min(1-a^2,b^2)\min(a^2,1-b^2)}{ab\sqrt{(1-a^2)(1-b^2)}}\le 1
\end{align}

From Theorem \ref{necc}, we have 
\begin{equation}
-\lambda_2(\tilde{P}_{UV})\le\lambda\le\lambda_2(\tilde{P}_{UV})
\end{equation}
Thus, from above, we have
\begin{equation}\label{bibo1}
-\min(\xi_2,\lambda_2(\tilde{P}_{UV})) \le\lambda\le \min(\xi_1,\lambda_2(\tilde{P}_{UV}))
\end{equation}

A sharper bound in this special case can be obtained as follows.
\begin{Theo}\label{bine}
If $X_1\longrightarrow U^n
\longrightarrow V^n \longrightarrow X_2$, and $(X_1, X_2, U^n, V^n)$ satisfies the above settings, then for sufficiently large $n$, 
\begin{equation}\label{bibo2}
-\min\left(\xi_2,\lambda_2(\tilde{P}_{UV})\frac{1+\xi_2}{2}\right)
\le
\lambda
\le
\min\left(\xi_1,\lambda_2(\tilde{P}_{UV})\frac{1+\xi_1}{2}\right)
\end{equation}
\end{Theo}
The proof of Theorem \ref{bine} is given in Appendix \ref{bicase}.

The bound in (\ref{bibo2}) is tighter than the one in (\ref{bibo1}) because $\xi_1\le 1$ and therefore $\frac{1+\xi_1}{2}\le 1$. A similar argument holds for the other side of the inequality as well.

In the above derivation, we provided two necessary conditions for the $n$-letter Markov chain $X_1\longrightarrow U^n
\longrightarrow V^n \longrightarrow X_2$, where $n\rightarrow\infty$, in this special case of binary random variables. In other words, we provided two outer bounds for $\lambda$, where the joint distributions $p(x_1,x_2,u^n,v^n)$ satisfy the $n$-letter Markov chain $X_1\longrightarrow U^n
\longrightarrow V^n \longrightarrow X_2$ with $n\rightarrow\infty$ and satisfy the fixed marginal distributions given in (\ref{marab1}) and (\ref{marab2}).

For reference, we give a sufficient condition for $X_1\rightarrow U^n\rightarrow V^n\rightarrow X_2$,
 or equivalently, an inner bound for $\lambda$ satisfying this $n$-letter Markov chain. This inner bound is obtained by noting that if $(X_1,X_2)$ satisfies $X_1\rightarrow U\rightarrow V\rightarrow X_2$, then it satisfies $X_1\rightarrow U^n\rightarrow V^n\rightarrow X_2$. In this case, using Theorem \ref{iff} we have
\begin{equation}
\lambda=\lambda_L\lambda_{2}(\tilde{P}_{UV})\lambda_R
\end{equation}
where $\lambda_L$ and $\lambda_R$ are such that
\begin{align}
\tilde{P}_{X_1U}&\triangleq\left[\begin{array}{c}a\\\sqrt{1-a^2}\end{array}\right]
\left[\begin{array}{cc}\frac{1}{\sqrt{2}}&\frac{1}{\sqrt{2}}\end{array}\right]+
\lambda_L\left[\begin{array}{c}\sqrt{1-a^2}\\-a\end{array}\right]\left[\begin{array}{cc}\frac{1}{\sqrt{2}}&-\frac{1}{\sqrt{2}}\end{array}\right]\ge 0\\
\tilde{P}_{VX_2}&
\triangleq\left[\begin{array}{c}\frac{1}{\sqrt{2}}\\\frac{1}{\sqrt{2}}\end{array}\right]
\left[\begin{array}{cc}b&\sqrt{1-b^2}\end{array}\right]+
\lambda_R
\left[\begin{array}{c}\frac{1}{\sqrt{2}}\\-\frac{1}{\sqrt{2}}\end{array}\right]\left[\begin{array}{cc}\sqrt{1-b^2}&-b\end{array}\right]\ge0
\end{align}
Due to the non-negativity of the matrices $\tilde{P}_{X_1U}$ and $\tilde{P}_{VX_2}$, we have 
\begin{align}
-\frac{\min(a^2, 1-a^2)}{a\sqrt{1-a^2}}&\le\lambda_L\le\frac{\min(a^2, 1-a^2)}{a\sqrt{1-a^2}}\\
-\frac{\min(b^2, 1-b^2)}{b\sqrt{1-b^2}}&\le\lambda_R\le\frac{\min(b^2, 1-b^2)}{b\sqrt{1-b^2}}
\end{align}
Thus, we have 
\begin{equation}\label{bibo3}
-\lambda_{2}(\tilde{P}_{UV})\xi_3\le
\lambda\le
\lambda_{2}(\tilde{P}_{UV})\xi_3
\end{equation}
where
\begin{align}
\xi_3&\triangleq\frac{\min(a^2,1-a^2)\min(b,1-b^2)}{ab\sqrt{(1-a^2)(1-b^2)}}
\end{align}
Then, combining (\ref{bibo1}), (\ref{bibo2}), and (\ref{bibo3}), we have the two outer bounds and one inner bound for $\lambda$ as follows
\begin{align}
\lambda_{2}(\tilde{P}_{UV})\xi_3&\le
\sup_{X_1\rightarrow U^n\rightarrow V^n\rightarrow X_2}\lambda\le
\min(\xi_1,\lambda_2(\tilde{P}_{UV})\frac{1+\xi_1}{2})\le
\min(\xi_1,\lambda_2(\tilde{P}_{UV}))
\\
-\min(\xi_2,\lambda_2(\tilde{P}_{UV}))&\le
-\min(\xi_2,\lambda_2(\tilde{P}_{UV})\frac{1+\xi_2}{2})
\le \inf_{X_1\rightarrow U^n\rightarrow V^n\rightarrow X_2}\lambda
\le-\lambda_{2}(\tilde{P}_{UV})\xi_3
\end{align} 

\begin{figure}
\centering
\includegraphics[width=6.5in]{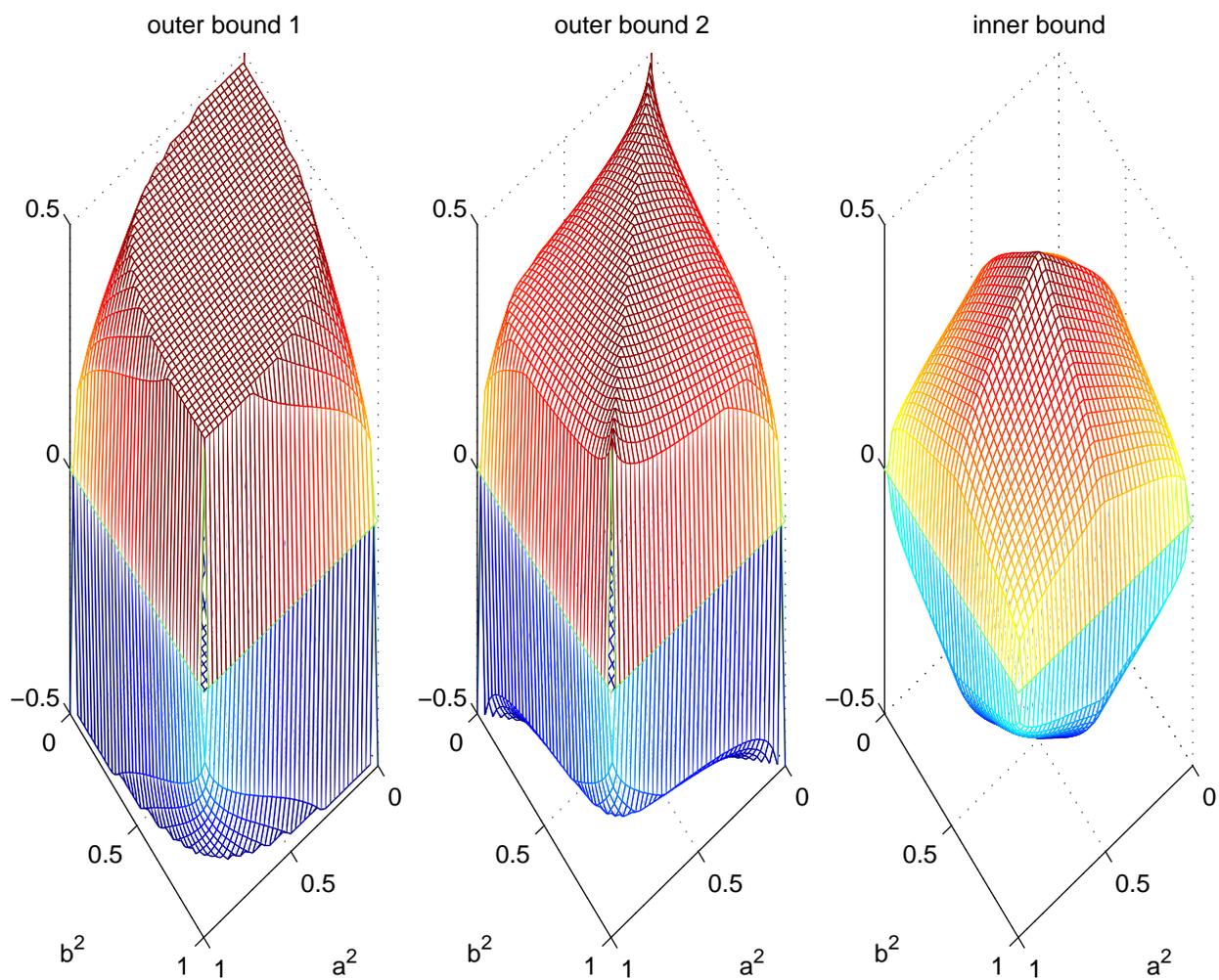}
\caption{(i) Outer bound 1, (ii) outer bound 2, and (iii) inner bound for $\lambda$.} \label{binai}
\end{figure}

We illustrate these three bounds with $\lambda_2(\tilde{P}_{UV})=0.5$ in Figure \ref{binai}.

\section{Proofs of Some Theorems}\label{proofs}
\subsection{Proof of Theorem \ref{app}}\label{proofs1}

To find $\underset{F(n, P_{X_1}),\; n=1,2,\dots}{\sup} \lambda_2(\tilde{P}_{X_1U^n})$,
we need to exhaust the sets $F(n, P_{X_1})$ with $n\ge 1$.
In the following, we show that it suffices to check only the asymptotic case. 

For any joint distribution $P_{X_1U^n}\in F(n, P_{X_1})$,
we attach an independent $U$, say $U_{n+1}$, to the existing $n$-sequence,
and get a new joint distribution $P_{X_1U^{n+1}}=P_{X_1U^n}\otimes p_{U}$,
where $p_{U}$ is the marginal distribution of $U$ in the vector form.
By arguments similar to those in Section \ref{condD},
we have that
$\lambda_i(\tilde{P}_{X_1U^{n+1}})=\lambda_i(\tilde{P}_{X_1U^{n}})$.
Therefore, for every $P_{X_1U^n}\in F(n, P_{X_1})$, there exists some $P_{X_1U^{n+1}}\in F(n+1, P_{X_1})$, 
such that $\lambda_i(\tilde{P}_{X_1U^{n+1}})=\lambda_i(\tilde{P}_{X_1U^{n}})$.
Thus,
\begin{equation}
\underset{F(n, P_{X_1})}{\sup} \lambda_2(\tilde{P}_{X_1U^n})\le
\underset{F(n+1, P_{X_1})}{\sup} \lambda_2(\tilde{P}_{X_1U^{n+1}})\label{mono}
\end{equation}
From (\ref{mono}),  we see that $\underset{F(n, P_{X_1})}{\sup} \lambda_2(\tilde{P}_{X_1U^n})$
is monotonically non-decreasing in $n$. We also note that $\lambda_{2}(\tilde{P}_{X_1U^n})$ is upper bounded by $1$
for all $n$, i.e.,~$\lambda_{2}(\tilde{P}_{X_1U^n})\le 1$.
 Therefore,
\begin{equation}
\underset{F(n, P_{X_1}),\; n=1,2,\dots}{\sup} \lambda_2(\tilde{P}_{X_1U^n})=
\lim_{n\rightarrow\infty}\underset{F(n, P_{X_1})}{\sup} \lambda_2(\tilde{P}_{X_1U^n})\label{lim}
\end{equation}
To complete the proof, we need the following lemma. 
\begin{Lem}\cite{Witsenhausen:1975} \label{decom}
 $\lambda_{2}(\tilde{P}_{XY})=1$ if and only if $P_{XY}$ decomposes.
By $P_{XY}$ decomposes, we mean that there exist sets $S_1\in\mathcal{X}$,
$S_2\in\mathcal{Y}$, such that $P(S_1)$, $P(\mathcal{X}-S_1)$, $P(S_2)$, $P(\mathcal{Y}-S_2)$
are positive, while $P((\mathcal{X}-S_1)\times S_2)=P(S_1\times (\mathcal{Y}-S_2))=0$.
\end{Lem}

In the following, we will show by construction that 
there exists a joint distribution that 
decomposes asymptotically.

For a given marginal distribution $P_{X_1}$, we arbitrarily choose a subset $S_1$ from
 the alphabet of $X_1$ with positive $P(S_1)$. We  find a set $S_2$ in the alphabet of $U^n$ such that
 $P(S_1)=P(S_2)$ if it is possible. Otherwise, we pick $S_2$ with positive $P(S_2)$ such that $|P(S_1)-P(S_2)|$ is minimized.
We denote $\mathcal{L}(n)$ to be the set of all subsets of the alphabet
 of $U^n$ and we also define $P_{\max}=\max Pr(s)$ for all $s\in \mathcal{U}$.
 Then, we have
\begin{equation}\label{upb}
\underset{S_2\subset\mathcal{L}(n)}{\min}|P(S_2)-P(S_1)|\le P_{\max}^n
\end{equation}

We construct a joint distribution for $X_1$ and $U^n$ as follows.
First, we construct the joint distribution $P^i$ corresponding to the case where 
$X_1$ and $U^n$ are independent.
Second, we rearrange the alphabets of $X_1$ and $U^n$ and group the sets
 $S_1$, $\mathcal{X}_1-S_1$, $S_2$ and $\mathcal{U}^n-S_2$ as follows
\begin{equation}
P^i=\left[\begin{array}{ll}P_{11}^i&P_{12}^i\\P_{21}^i&P_{22}^i\end{array}\right]\label{Pi}
\end{equation}
where $P_{11}^i$, $P_{12}^i$, $P_{21}^i$, $P_{22}^i$    correspond to the sets $S_1\times S_2$,
$S_1\times (\mathcal{U}^n-S_2)$,
$(\mathcal{X}_1-S_1)\times S_2$,
$(\mathcal{X}_1-S_1)\times (\mathcal{U}^n-S_2)$, respectively.
Here, we assume that $P(S_2)\ge P(S_1)$.
Then, we scale these four sub-matrices as
$P_{11}=\frac{P_{11}^iP(S_1)}{P(S_1)P(S_2)}$,
$P_{12}=0$,
$P_{21}=\frac{P_{21}^i(P(S_2)-P(S_1))}{(1-P(S_1))P(S_2)}$,
$P_{22}=\frac{P_{22}^i(1-P(S_2))}{(1-P(S_1))(1-P(S_2))}$,
and let
\begin{equation}
P=\left[\begin{array}{ll}P_{11}&0\\P_{21}&P_{22}\end{array}\right]
\end{equation}
We note that $P$ is a joint distribution for $X_1$ and $U^n$ with the given marginal
distributions.
Next, we move the mass in the sub-matrix $P_{21}$ to $P_{11}$, which yields
\begin{equation}
P'\!\triangleq\!\left[\begin{array}{ll}P_{11}'&0\\0&P_{22}\end{array}\right]
\!\!=P+E=\!\!\left[\begin{array}{ll}P_{11}&0\\P_{21}&P_{22}\end{array}\right]
+\left[\begin{array}{ll}E_{11}&0\\-E_{21}&0\end{array}\right]\label{Pp}
\end{equation}
where
$E_{21}\triangleq P_{21}$, $E_{11}\triangleq\frac{P_{11}^i(P(S_2)-P(S_1))}{P(S_1)P(S_2)}$,
and $P_{11}'=\frac{P_{11}P(S_2)}{P(S_1)}$.
We denote $P'_{X_1}$ and $P'_{U^n}$ as the marginal distributions of $P'$.
We note that $P'_{U^n}=P_{U^n}$ and $P'_{X_1}=P_{X_1}M$ where $M$ is a scaling
diagonal matrix. The elements in the set $S_1$ are scaled up by a factor of $\frac{P(S_2)}{P(S_1)}$,
and those in the set $\mathcal{X}_1-S_1$ are scaled down by a factor of $\frac{1-P(S_2)}{1-P(S_1)}$.
Then,
\begin{align}
\tilde{P}'
&=M^{-\frac{1}{2}}\tilde{P}+M^{-\frac{1}{2}}P_{X_1}^{-\frac{1}{2}}EP_{U^n}^{-\frac{1}{2}}
\end{align}
We will need the following lemmas in the remainder of our derivations.
Lemma \ref{times} can be proved using techniques similar to those in the proof of Lemma \ref{sum} \cite{Stewart:1993}.
\begin{Lem}\cite{Stewart:1993}\label{sum}
If $A'=A+E$, then $|\lambda_{i}(A')-\lambda_{i}(A)|\le||E||_2$, where $||E||_2$ is the spectral norm of $E$.
\end{Lem}
\begin{Lem}\label{times}
If $A'=MA$, where $M$ is an invertible matrix,
then $||M^{-1}||_2^{-1}\le\lambda_{i}(A')/\lambda_{i}(A)\le||M||_2$.
\end{Lem}

Since $P'$ decomposes,  using Lemma \ref{decom}, we conclude that  $\lambda_2(\tilde{P}')=1$.
We upper bound  $||P_{X_1}^{-\frac{1}{2}}EP_{U^n}^{-\frac{1}{2}}||_2$ as follows,
\begin{equation}
||P_{X_1}^{-\frac{1}{2}}EP_{U^n}^{-\frac{1}{2}}||_2
\le
||P_{X_1}^{-\frac{1}{2}}EP_{U^n}^{-\frac{1}{2}}||_F
\label{22f}
\end{equation}
where $||\cdot||_F$ is the Frobenius norm.
Combining (\ref{Pi}) and (\ref{Pp}), we have
\begin{align}
||P_{X_1}^{-\frac{1}{2}}&EP_{U^n}^{-\frac{1}{2}}||_F
\le\frac{(P(S_2)-P(S_1))}{P_1'P(S_2)}
||P_{X_1}^{-\frac{1}{2}}P^iP_{U^n}^{-\frac{1}{2}}||_F\label{fup}
\end{align}
where $P_1'\triangleq\min(P(S_1),1-P(S_1))$. Since $P^i$ corresponds to
the independent case, we have $||P_{X_1}^{-\frac{1}{2}}P^iP_{U^n}^{-\frac{1}{2}}||_F=1$
from (\ref{fun}). Then, from (\ref{upb}), (\ref{22f}) and (\ref{fup}),  we obtain
\begin{equation}
||P_{X_1}^{-\frac{1}{2}}EP_{U^n}^{-\frac{1}{2}}||_2\le c_1P_{\max}^n
\end{equation}
where $c_1\triangleq\frac{1}{P_1'P(S_2)}$.

From Lemma \ref{pro}, we have
\begin{align}
||M^{-\frac{1}{2}}&P_{X_1}^{-\frac{1}{2}}EP_{U^n}^{-\frac{1}{2}}||_2=
|\lambda_{1}(M^{-\frac{1}{2}}P_{X_1}^{-\frac{1}{2}}EP_{U^n}^{-\frac{1}{2}})|\le\left(\frac{1-P(S_1)}{1-P(S_2)}\right)^{\frac{1}{2}}c_1P_{\max}^n\triangleq c_2P_{\max}^n
\end{align}
From Lemma \ref{sum}, we have
\begin{equation}
1-c_2P_{\max}^{n}\le\lambda_2(M^{-\frac{1}{2}}\tilde{P})\le 1+c_2P_{\max}^{n}
\end{equation}
We upper bound $||M^{\frac{1}{2}}||_2$ as follows
\begin{align}
||M^{\frac{1}{2}}||_2=&\sqrt{\frac{P(S_2)}{P(S_1)}}
\le1+\sqrt{\frac{P(S_2)-P(S_1)}{P(S_1)}}
\le1+\frac{P_{\max}^{n/2}}{\sqrt{P(S_1)}}\triangleq1+c_3P_{\max}^{n/2}
\end{align}
Similarly,
$
||M^{-\frac{1}{2}}||_2^{-1}\ge1-c_4P_{\max}^{n/2}
$. 
From Lemma \ref{times}, we have
\begin{equation}
(1-c_4P_{\max}^{n/2})\le\frac{\lambda_2(\tilde{P})}{\lambda_2(M^{-\frac{1}{2}}\tilde{P})}
\le(1+c_3P_{\max}^{n/2})
\end{equation}
Since $P$ is a joint distribution matrix, from Theorem \ref{iff}, we know that $\lambda_2(\tilde{P})\le 1$.
Therefore, we have
\begin{align}
(1-c_4P_{\max}^{n/2})(1-c_2P_{\max}^{n})&\le\lambda_2(\tilde{P})\le1
\end{align}
When $P_{\max}<1$, corresponding to the non-trivial case,
$\lim_{n\rightarrow\infty} P_{\max}^{n/2}= 0$, 
and using (\ref{lim}), (\ref{mapping}) follows.

The case $P(S_2)< P(S_1)$
can be proved similarly. 
$\blacksquare$ 

\subsection{Proof of Theorem \ref{bine}} \label{bicase}
From (\ref{SVDbi}), we know
\begin{align}
\tilde{P}_{UV}
&=\left[\begin{array}{c}\frac{1}{\sqrt{2}}\\\frac{1}{\sqrt{2}}\end{array}\right]
\left[\begin{array}{cc}\frac{1}{\sqrt{2}}&\frac{1}{\sqrt{2}}\end{array}\right]+
\lambda_2(\tilde{P}_{UV})\begin{bmatrix}\frac{1}{\sqrt{2}}\\-\frac{1}{\sqrt{2}}\end{bmatrix}\begin{bmatrix}\frac{1}{\sqrt{2}}&-\frac{1}{\sqrt{2}}\end{bmatrix}
\end{align} 
From (\ref{SVDPN}), we know 
\begin{equation}
\tilde{P}_{U^nV^n}=\tilde{P}_{UV}^{\otimes n}=\frac{1}{2^n}
\left[\begin{array}{c}1\\\vdots\\1\end{array}\right]
\left[\begin{array}{ccc}1&\cdots&1\end{array}\right]+
\sum_{i=2}^{2^n}\lambda_2(\tilde{P}_{UV})^{l_i}\bm{\mu}_{i}(\tilde{P}_{U^nV^n})\bm{\nu}_i^T(\tilde{P}_{U^nV^n})
\end{equation} 
where $l_i\in\{1,2,\dots,n\}$, for $i=2,\dots, 2^n$. Due to the symmetric structure of $\tilde{P}_{U^nV^n}$, we have
\begin{align}
\bm{\mu}_{i}(\tilde{P}_{U^nV^n})&=\bm{\nu}_i(\tilde{P}_{U^nV^n}),\qquad i=2,\dots, 2^n
\end{align}
We also have
\begin{equation}
\tilde{P}_{X_1U^n}=\frac{1}{2^{n/2}}\left[\begin{array}{c}a\\\sqrt{1-a^2}\end{array}\right]
\left[\begin{array}{ccc}1&\cdots&1\end{array}\right]+\left[\begin{array}{c}\sqrt{1-a^2}\\-a\end{array}\right]\mathbf{c}^T
\end{equation}
where $\mathbf{c}$ is the product of the second singular value and the second right singular vector of $\tilde{P}_{X_1U^n}$.
Similarly,
\begin{equation}
\tilde{P}_{V^nX_2}=\frac{1}{2^{n/2}}
\left[\begin{array}{c}1\\\vdots\\1\end{array}\right]
\left[\begin{array}{cc}b&\sqrt{1-b^2}\end{array}\right]
+\mathbf{d}\left[\begin{array}{cc}\sqrt{1-b^2}&-b\end{array}\right]
\end{equation}
From (\ref{righ}), we know that
\begin{align}
\tilde{P}_{X_1X_2}=&\tilde{P}_{X_1U^n}\tilde{P}_{U^nV^n}\tilde{P}_{V^nX_2}\nonumber\\
=&\left[\begin{array}{c}a\\\sqrt{1-a^2}\end{array}\right]
\left[\begin{array}{cc}b&\sqrt{1-b^2}\end{array}\right]+\nonumber\\
&\left[\begin{array}{c}\sqrt{1-a^2}\\-a\end{array}\right]
\mathbf{c}^T\left(\sum_{i=2}^{2^n}\lambda_2(\tilde{P}_{UV})^{l_i}
\bm{\mu}_{i}(\tilde{P}_{U^nV^n})\bm{\nu}_i(\tilde{P}_{U^nV^n})\right)
\mathbf{d}\left[\begin{array}{cc}\sqrt{1-b^2}&-b\end{array}\right]\label{case11}
\end{align} 
Thus, we conclude that,
\begin{align}
\lambda&=\mathbf{c}^T\left(\sum_{i=2}^{2^n}\lambda_2(\tilde{P}_{UV})^{l_i}
\bm{\mu}_{i}(\tilde{P}_{U^nV^n})\bm{\nu}_i^T(\tilde{P}_{U^nV^n})\right)
\mathbf{d}
\end{align}
Consider the following optimization problem,
\begin{equation}
\max\quad \lambda=\max_{\mathbf{c},\mathbf{d}} \quad
\mathbf{c}^T\left(\sum_{i=2}^{2^n}\lambda_2(\tilde{P}_{UV})^{l_i}
\bm{\mu}_{i}(\tilde{P}_{U^nV^n})\bm{\nu}_i^T(\tilde{P}_{U^nV^n})\right)
\mathbf{d}
\end{equation}
We define 
\begin{align}
\gamma_i&\triangleq \mathbf{c}^T\bm{\mu}_{i}(\tilde{P}_{U^nV^n}),\quad i=2,\dots,2^n\\
\delta_i&\triangleq \mathbf{d}^T\bm{\nu}_{i}(\tilde{P}_{U^nV^n}),\quad i=2,\dots,2^n
\end{align}
Then,
\begin{equation}
\lambda=\sum_{i=2}^{2^n}\lambda_2(\tilde{P}_{UV})^{l_i}\gamma_i\delta_i
\end{equation}
We partition the set $\{2,\dots,2^n\}$ into two disjoint subsets, $\mathcal{L}^+$ and $\mathcal{L}^-$, such that 
\begin{equation}\label{s+s-}
i\in\left\{\begin{array}{cl}\mathcal{L}^+&\text{ if }\gamma_i\delta_i\ge 0\\
\mathcal{L}^-&\text{ if }\gamma_i\delta_i< 0
\end{array}\right. \quad  i=1,\dots,2^n
\end{equation}
Hence,
\begin{align}
\lambda
&=\sum_{i\in \mathcal{S}^+}\lambda_2(\tilde{P}_{UV})^{l_i}\gamma_i\delta_i
+\sum_{i\in \mathcal{S}^-}\lambda_2(\tilde{P}_{UV})^{l_i}\gamma_i\delta_i\nonumber\\
&\overset{1}{\le} \lambda_{2}(\tilde{P}_{UV}) \sum_{i\in \mathcal{S}^+}\gamma_i\delta_i\nonumber\\
&\overset{2}{\le} \frac{\lambda_{2}(\tilde{P}_{UV})}{4} \sum_{i\in \mathcal{S}^+}(\gamma_i+\delta_i)^2\nonumber\\
&\overset{3}{\le} \frac{\lambda_{2}(\tilde{P}_{UV})}{4} \sum_{i=2}^{2^n}(\gamma_i+\delta_i)^2\nonumber\\
&\overset{4}{=} \frac{\lambda_{2}(\tilde{P}_{UV})}{4} (\mathbf{c}+\mathbf{d})^T(\mathbf{c}+\mathbf{d})\nonumber\\
&= \frac{\lambda_{2}(\tilde{P}_{UV})}{2}\left(\frac{\mathbf{c}^T\mathbf{c}+\mathbf{d}^T\mathbf{d}}{2}+\mathbf{c}^T\mathbf{d}\right)\nonumber\\
&\overset{5}{\le} \frac{\lambda_{2}(\tilde{P}_{UV})}{2}(1+\mathbf{c}^T\mathbf{d})
\end{align}
where
\begin{enumerate}
\item because of the definition of $\mathcal{L}^+$ and $\mathcal{L}^-$ in (\ref{s+s-}) 
and  $0\le\lambda_{2}(\tilde{P}_{UV})\le1$;
\item because for non-negative $ \gamma_i\delta_i$,
\begin{equation}
(\gamma_i-\delta_i)^2=
\gamma_i^2+\delta_i^2-2\gamma_i\delta_i\ge 0
\end{equation}
Hence, by adding $4\gamma_i\delta_i$ to both sides of the above inequality, we have
\begin{equation}
(\gamma_i+\delta_i)^2\ge 4\gamma_i\delta_i
\end{equation}
\item due to the fact that $(\gamma_i+\delta_i)^2$ is non-negative for $i\in \mathcal{L}^-$;
\item comes from the following derivation
\begin{align}
\sum_{i=2}^{2^n}(\gamma_i+\delta_i)^2
=&\sum_{i=2}^{2^n}\bigg(\mathbf{c}^T\bm{\mu}_{i}(\tilde{P}_{U^nV^n})+
\mathbf{d}^T\bm{\nu}_{i}(\tilde{P}_{U^nV^n})\bigg)^2\nonumber\\
=&\sum_{i=2}^{2^n}\bigg((\mathbf{c}+\mathbf{d})^T\bm{\mu}_{i}(\tilde{P}_{U^nV^n})\bigg)^2\nonumber\\
\overset{(a)}{=}&\sum_{i=1}^{2^n}\bigg((\mathbf{c}+\mathbf{d})^T\bm{\mu}_{i}(\tilde{P}_{U^nV^n})\bigg)^2\nonumber\\
=&(\mathbf{c}+\mathbf{d})^TMM^T(\mathbf{c}+\mathbf{d})\nonumber\\
\overset{(b)}{=}&(\mathbf{c}+\mathbf{d})^T(\mathbf{c}+\mathbf{d})
\end{align}
where 
\begin{enumerate}
\item
 because both the vectors $\mathbf{c}$ and $\mathbf{d}$ are within 
the subspace spanned by singular vectors
$[\bm{\mu}_{2}(\tilde{P}_{U^nV^n}),
\cdots,\bm{\mu}_{2^n}(\tilde{P}_{U^nV^n})]$, thus
\begin{equation}
(\mathbf{c}+\mathbf{d})^T\bm{\mu}_{1}(\tilde{P}_{U^nV^n})=0
\end{equation}
\item because
\begin{equation}
MM^T=I
\end{equation}
\end{enumerate}
\item because $\mathbf{c}^T\mathbf{c}=\lambda_{2}(\tilde{P}_{X_1U^n})^2$ 
and $\mathbf{d}^T\mathbf{d}=\lambda_{2}(\tilde{P}_{V^nX_2})^2$ and
from Theorem \ref{iff}, we know that the square of $\lambda_{2}$ is  less than or equal to $1$.
\end{enumerate}
From the above discussion, we conclude that
\begin{equation}
\max\quad \lambda
\le \max_{\mathbf{c},\mathbf{d}}\quad \frac{\lambda_{2}(\tilde{P}_{UV})}{2}(1+\mathbf{c}^T\mathbf{d})
\end{equation}
Thus, we can upper bound $\lambda$ by $\max_{\mathbf{c},\mathbf{d}}
\frac{\lambda_{2}(\tilde{P}_{UV})}{2}(1+\mathbf{c}^T\mathbf{d})$. 

\begin{figure}
\centering
\includegraphics[width=3.5in]{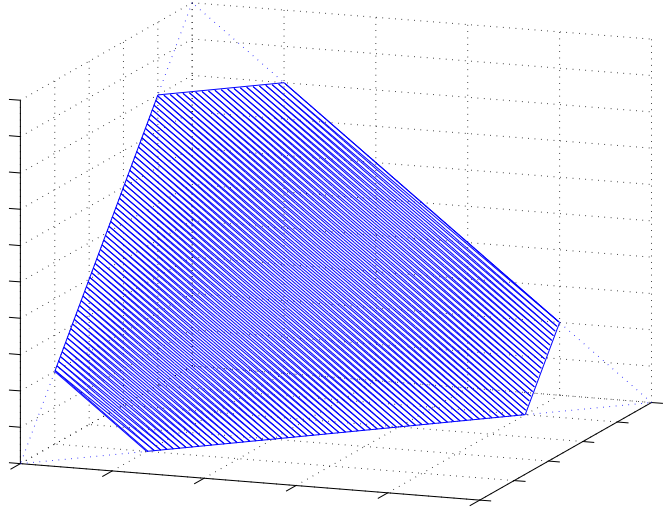}
\caption{Subset of simplex satisfying (\ref{bound}).}\label{sim}
\end{figure}

From (\ref{def}), we know that $\tilde{P}_{X_1U^n}$ is a non-negative matrix, i.e.,~
\begin{equation}
\tilde{P}_{X_1U^n}
=\left[\begin{array}{c}\frac{1}{2^{n/2}}a\mathbf{e}^T+\sqrt{1-a^2}\mathbf{c}^T\\
\frac{1}{2^{n/2}}\sqrt{1-a^2}\mathbf{e}^T-a\mathbf{c}^T\end{array}\right]
\ge \mathbf{0}
\end{equation}
where $\mathbf{e}$ is defined as a vector where all its elements are equal to $1$, and for matrix $A$ and $B$, by $\mathbf{A}\ge\mathbf{B}$, we mean all the entries of the matrix $\mathbf{A}-\mathbf{B}$ are non-negative. 
This property implies that
\begin{equation}\label{bound}
\frac{1}{2^{n/2}}\frac{1}{a\sqrt{1-a^2}}\mathbf{e}\ge\bar{\mathbf{c}}\triangleq\frac{1}{2^{n/2}}\frac{a}{\sqrt{1-a^2}}\mathbf{e}+\mathbf{c}\ge\mathbf{0}
\end{equation}
We know that $\mathbf{c}$ is orthogonal to $\mathbf{e}$, i.e.,
\begin{equation}\label{ortho}
\mathbf{c}^T\mathbf{e}=\sum_{i=1}^{2^n}c_i=0
\end{equation}
Hence,  we see that
 the vector $\bar{\mathbf{c}}$ is on the hyperplane that contains the point
 $\frac{1}{2^{n/2}}\frac{a}{\sqrt{1-a^2}}\mathbf{e}$ and
 is orthogonal to the vector $\mathbf{e}$. 
 On the other hand, (\ref{bound}) shows that each coordinate of $\bar{\mathbf{c}}$
 is non-negative and less than or equal to $\frac{1}{2^{n/2}}\frac{1}{a\sqrt{1-a^2}}$.
 Thus, the vector $\bar{\mathbf{c}}$ lies on a subset of simplex.
 See Figure \ref{sim} for a three-dimension illustration.

By a symmetric argument, we have
\begin{equation}
\frac{1}{2^{n/2}}\frac{1}{b\sqrt{1-b^2}}\mathbf{e}
\ge\bar{\mathbf{d}}\triangleq
\frac{1}{2^{n/2}}\frac{b}{\sqrt{1-b^2}}\mathbf{e}+
\mathbf{d}\ge\mathbf{0}
\end{equation} 
Since $\bar{\mathbf{c}}\triangleq\frac{1}{2^{n/2}}\frac{a}{\sqrt{1-a^2}}\mathbf{e}+\mathbf{c}$ 
and $\bar{\mathbf{d}}\triangleq
\frac{1}{2^{n/2}}\frac{b}{\sqrt{1-b^2}}\mathbf{e}+\mathbf{d}$,
\begin{align}
\bar{\mathbf{c}}^T\bar{\mathbf{d}}=&
\left(\frac{1}{2^{n/2}}\frac{a}{\sqrt{1-a^2}}\mathbf{e}+\mathbf{c}\right)^T
\left(\frac{1}{2^{n/2}}\frac{b}{\sqrt{1-b^2}}\mathbf{e}+\mathbf{d}\right)\nonumber\\
=&\frac{ab}{\sqrt{(1-a^2)(1-b^2)}}+
\frac{1}{2^{n/2}}\frac{a}{\sqrt{1-a^2}}\mathbf{e}^T\mathbf{d}+
\frac{1}{2^{n/2}}\frac{b}{\sqrt{1-b^2}}\mathbf{e}^T\mathbf{c}+
\mathbf{c}^T\mathbf{d}\nonumber\\
=&\frac{ab}{\sqrt{(1-a^2)(1-b^2)}}+\mathbf{c}^T\mathbf{d}
\end{align} Then, 
\begin{equation}
\max_{\mathbf{c},\mathbf{d}}\quad \mathbf{c}^T\mathbf{d}=
\max_{\bar{\mathbf{c}},\bar{\mathbf{d}}}\quad \bar{\mathbf{c}}^T\bar{\mathbf{d}}
-\frac{ab}{\sqrt{(1-a^2)(1-b^2)}}
\end{equation} The feasible sets of $\bar{\mathbf{c}}$ and $\bar{\mathbf{d}}$ are defined as follows,
\begin{align}
\mathcal{C}\triangleq&\left\{\mathbf{x}:
\frac{1}{2^{n/2}}\frac{1}{a\sqrt{1-a^2}}\mathbf{e}
\ge\mathbf{x}\ge\mathbf{0}\quad\text{and}\quad
\mathbf{e}^T\mathbf{x}=2^{n/2}\frac{a}{\sqrt{1-a^2}}
\right\}\\
\mathcal{D}\triangleq&\left\{\mathbf{x}:
\frac{1}{2^{n/2}}\frac{1}{b\sqrt{1-b^2}}\mathbf{e}
\ge\mathbf{x}\ge\mathbf{0}\quad\text{and}\quad
\mathbf{e}^T\mathbf{x}=2^{n/2}\frac{b}{\sqrt{1-b^2}}
\right\}
\end{align} 
Consider the following optimization problem
\begin{equation}
\max_{\bar{\mathbf{c}}\in\mathcal{C}, \bar{\mathbf{d}}\in\mathcal{D}}\quad \bar{\mathbf{c}}^T\bar{\mathbf{d}}
\end{equation}
In the following, we will show that there exist $\mathcal{C}'\subseteq\mathcal{C}$ and $\mathcal{D}'\subseteq\mathcal{D}$ such that 
\begin{equation}
\max_{\bar{\mathbf{c}}\in\mathcal{C}, \bar{\mathbf{d}}\in\mathcal{D}}\quad \bar{\mathbf{c}}^T\bar{\mathbf{d}}
=\max_{\bar{\mathbf{c}}\in\mathcal{C}', \bar{\mathbf{d}}\in\mathcal{D}'}\quad \bar{\mathbf{c}}^T\bar{\mathbf{d}}
\end{equation}
If we assume that 
\begin{align}
\max_{\bar{\mathbf{c}}\in\mathcal{C}}\quad \bar{\mathbf{c}}^T\bar{\mathbf{d}}&=\max_{\bar{\mathbf{c}}\in\mathcal{C}'}\quad \bar{\mathbf{c}}^T\bar{\mathbf{d}}\qquad \forall\bar{\mathbf{d}}\in\mathcal{D}\label{maxc}\\
\max_{\bar{\mathbf{d}}\in\mathcal{D}}\quad \bar{\mathbf{c}}^T\bar{\mathbf{d}}&=\max_{\bar{\mathbf{d}}\in\mathcal{D}'}\quad \bar{\mathbf{c}}^T\bar{\mathbf{d}}\qquad \forall\bar{\mathbf{c}}\in\mathcal{C}\label{maxd}
\end{align}
and we also assume that the set $\mathcal{C}'$ ($\mathcal{D}'$ respectively) does not depend on the value of $\bar{\mathbf{d}}$ ($\bar{\mathbf{c}}$), then we have
\begin{align}
\max_{\bar{\mathbf{c}}\in\mathcal{C}, \bar{\mathbf{d}}\in\mathcal{D}}\quad \bar{\mathbf{c}}^T\bar{\mathbf{d}}
&=\max_{\bar{\mathbf{c}}\in\mathcal{C}}\quad \max_{\bar{\mathbf{d}}\in\mathcal{D}}\quad \bar{\mathbf{c}}^T\bar{\mathbf{d}}\nonumber\\
&\overset{1}{=}\max_{\bar{\mathbf{c}}\in\mathcal{C}}\quad \max_{\bar{\mathbf{d}}\in\mathcal{D}'}\quad \bar{\mathbf{c}}^T\bar{\mathbf{d}}\nonumber\\
&\overset{2}{=}\max_{\bar{\mathbf{d}}\in\mathcal{D}'}\quad \max_{\bar{\mathbf{c}}\in\mathcal{C}}\quad \bar{\mathbf{c}}^T\bar{\mathbf{d}}\nonumber\\
&\overset{3}{=}\max_{\bar{\mathbf{d}}\in\mathcal{D}'}\quad \max_{\bar{\mathbf{c}}\in\mathcal{C}'}\quad \bar{\mathbf{c}}^T\bar{\mathbf{d}}\nonumber\\
&=\max_{\bar{\mathbf{c}}\in\mathcal{C}', \bar{\mathbf{d}}\in\mathcal{D}'}\quad \bar{\mathbf{c}}^T\bar{\mathbf{d}}
\end{align}
where
\begin{enumerate}
\item because of (\ref{maxd});
\item because we assume that the set $\mathcal{D}'$ does not depend on the value of $\bar{\mathbf{c}}$;
\item because of (\ref{maxc}).
\end{enumerate}
Now we need to show our assumptions, (\ref{maxc}) and (\ref{maxd}), are valid, for which we need the following lemma.
\begin{Lem}\label{extreme}\cite[p. 722]{Bertsekas:1999}
Let $\mathcal{C}$ be a convex subset of $\mathbb{R}^n$, and let $\mathcal{C}^{\ast}$ be the set 
of minima of a concave function $f: \mathcal{C}\longmapsto \mathbb{R}$ over $\mathcal{C}$. 
If $\mathcal{C}$ is closed and contains at least one extreme point, and $\mathcal{C}^{\ast}$ is
nonempty, then $\mathcal{C}^{\ast}$ contains some extreme point of $\mathcal{C}$.
\end{Lem}
Here the extreme point is defined as follows:
\begin{Def}\cite[p. 721]{Bertsekas:1999}
A vector $\mathbf{x}$ is said to be an \emph{extreme point} of a convex set $\mathcal{C}$ if $\mathbf{x}$ belongs to $\mathcal{C}$
and there do not exist vectors $\mathbf{y}\in\mathcal{C}$ and $\mathbf{z}\in\mathcal{C}$, with $\mathbf{y}\ne \mathbf{x}$ and $\mathbf{z}\ne \mathbf{x}$, and a scalar $\alpha\in(0,1)$
such that $\mathbf{x}=\alpha \mathbf{y}+(1-\alpha)\mathbf{z}$. An equivalent definition is that $\mathbf{x}$ cannot be expressed as a convex combination of some
vectors of $\mathcal{C}$, all of which are different from $\mathbf{x}$.
\end{Def}
Thus, if we assume
\begin{align}
\mathcal{C}'&\triangleq\{\text{extreme points of }\mathcal{C}\}\\
\mathcal{D}'&\triangleq\{\text{extreme points of }\mathcal{D}\}
\end{align}
(\ref{maxc}) and (\ref{maxd}) will be satisfied. 
We observe that the set $\mathcal{C}'$ (respectively, the set $\mathcal{D}'$), which consists of all the extreme points in the set $\mathcal{C}$ (in the set $\mathcal{D}$ ), does not depend on the value of $\bar{\mathbf{d}}$ ($\bar{\mathbf{c}}$).

Next, we  determine the extreme point set $\mathcal{C}'$  in the following lemma.
\begin{Lem}
The set $\mathcal{C}'$ consists of all the vectors, each of which contains $2^n a^2$ non-zero entries with value
 $\frac{1}{2^{n/2}}\frac{1}{a\sqrt{1-a^2}}$, when $n$ is sufficiently large.
\end{Lem}
\begin{proof}
We define the set $\mathcal{C}''$ as the set
where each element contains $2^n a^2$ non-zero entries equal to
 $\frac{1}{2^{n/2}}\frac{1}{a\sqrt{1-a^2}}$.
 It is easy to see that every vector in $\mathcal{C}''$ is within the set $\mathcal{C}$.
 We need to show that any vector in the set $\mathcal{C}$ is a convex combination of some
 vectors in $\mathcal{C}''$. This can be proven by induction. It is easy to see that, if a vector such that
$2^n-1$ out of $2^n$ entries take values from $\{0, \frac{1}{2^{n/2}}\frac{1}{a\sqrt{1-a^2}}\}$,
the last entry will  converge to $0$, 
when $n$ goes to infinity. Let $\mathbf{s}\in\mathcal{C}$ such that $l$ out of $2^n$ entries  take
 values in $(0, \frac{1}{2^{n/2}}\frac{1}{a\sqrt{1-a^2}})$. Then, we choose any $2$ out
 of these $l$ entries, which are equal to $\alpha$ and $\beta$, respectively. 
 If $\alpha+\beta\le \frac{1}{2^{n/2}}\frac{1}{a\sqrt{1-a^2}}$, then
 \begin{align}
 &\begin{bmatrix}\cdots&\alpha&\cdots&\beta&\cdots\end{bmatrix}\nonumber\\
 &\qquad\qquad=
  \frac{\beta}{\alpha+\beta}\begin{bmatrix}\cdots&0&\cdots&\alpha+\beta&\cdots\end{bmatrix} +
\frac{\alpha}{\alpha+\beta}   \begin{bmatrix}\cdots&\alpha+\beta&\cdots&0&\cdots\end{bmatrix}
 \end{align}
 If $\alpha+\beta\ge \frac{1}{2^{n/2}}\frac{1}{a\sqrt{1-a^2}}$, then
  \begin{align}
 &\begin{bmatrix}\cdots&\alpha&\cdots&\beta&\cdots\end{bmatrix}\nonumber\\
  &\qquad\qquad=\frac{\frac{1}{2^{n/2}}\frac{1}{a\sqrt{1-a^2}}-\beta}
  {\frac{2}{2^{n/2}}\frac{1}{a\sqrt{1-a^2}}-\alpha-\beta}
  \begin{bmatrix}\cdots&\frac{1}{2^{n/2}}\frac{1}{a\sqrt{1-a^2}}&\cdots&
  \alpha+\beta-\frac{1}{2^{n/2}}\frac{1}{a\sqrt{1-a^2}}&\cdots\end{bmatrix} \nonumber\\
  &\qquad\qquad\quad\;+
\frac{\frac{1}{2^{n/2}}\frac{1}{a\sqrt{1-a^2}}-\alpha}
{\frac{2}{2^{n/2}}\frac{1}{a\sqrt{1-a^2}}-\alpha-\beta}   
\begin{bmatrix}\cdots&\alpha+\beta-\frac{1}{2^{n/2}}\frac{1}{a\sqrt{1-a^2}}&
\cdots&\frac{1}{2^{n/2}}\frac{1}{a\sqrt{1-a^2}}&\cdots\end{bmatrix}
 \end{align}
which means that $\mathbf{s}$ can be expressed as a convex combination of two vectors. These two vectors
belong to set $\mathcal{C}$ and both of them have $l-1$ out of $2^n$ entries takes value in 
$(0, \frac{1}{2^{n/2}}\frac{1}{a\sqrt{1-a^2}})$. By induction, we can show that 
every vector in set $\mathcal{C}$ can be expressed as a convex combination of some vectors
in $\mathcal{C}''$.
 On the other hand, 
it is easy to see that any vector $\mathbf{s}$ in $\mathcal{C}''$ cannot be expressed as a convex combination of
some vectors in the set $\mathcal{C}$ other than $\mathbf{s}$ itself. Thus we conclude 
that $\mathcal{C}'=\mathcal{C}''$. 
\end{proof}

Similarly, the set $\mathcal{D}'$ consists all the  vectors, each of which contains $2^n b^2$ non-zero entries with value $\frac{1}{2^{n/2}}\frac{1}{b\sqrt{1-b^2}}$.
Then,
\begin{equation}
\max_{\bar{\mathbf{c}}\in\mathcal{C}, \bar{\mathbf{d}}\in\mathcal{D}}\quad \bar{\mathbf{c}}^T\bar{\mathbf{d}}
=\max_{\bar{\mathbf{c}}\in\mathcal{C}', \bar{\mathbf{d}}\in\mathcal{D}'}\quad \bar{\mathbf{c}}^T\bar{\mathbf{d}}
=\min(a^2,b^2)\frac{1}{a\sqrt{1-a^2}}\frac{1}{b\sqrt{1-b^2}}
\end{equation}
and,
\begin{align}
\max_{\mathbf{c},\mathbf{d}}\quad\mathbf{c}^T\mathbf{d}=&
\max_{\bar{\mathbf{c}}\in\mathcal{C}, \bar{\mathbf{d}}\in\mathcal{D}}\quad \bar{\mathbf{c}}^T\bar{\mathbf{d}}-
\frac{ab}{\sqrt{(1-a^2)(1-b^2)}}\nonumber\\
=&\min(a^2,b^2)\frac{1}{ab\sqrt{(1-a^2)(1-b^2)}}-\frac{ab}{\sqrt{(1-a^2)(1-b^2)}}\nonumber\\
=&\min(a^2,b^2)\min(1-a^2,1-b^2)\frac{1}{ab\sqrt{(1-a^2)(1-b^2)}}
\end{align}
Hence,
\begin{equation}
\lambda\le\lambda_{2}(\tilde{P}_{UV})\frac{1+\mathbf{c}^T\mathbf{d}}{2}\le 
\lambda_{2}(\tilde{P}_{UV})\frac{1+\frac{\min(a^2,b^2)\min(1-a^2,1-b^2)}{ab\sqrt{(1-a^2)(1-b^2)}}}{2}\label{cs1}
\end{equation}
The lower bound of $\lambda$ 
can be derived in a similar manner. %
We   rewrite (\ref{case11}) in the following form
\begin{align}
\tilde{P}_{X_1X_2}
&=\left[\begin{array}{c}a\\\sqrt{1-a^2}\end{array}\right]
\left[\begin{array}{cc}b&\sqrt{1-b^2}\end{array}\right]+(-\lambda)
\left[\begin{array}{c}\sqrt{1-a^2}\\-a\end{array}\right]
\left[\begin{array}{cc}-\sqrt{1-b^2}&b\end{array}\right]\label{case3}
\end{align}
By the same arguments as above, we obtain
\begin{align}
-\lambda&\le
\lambda_{2}(\tilde{P}_{UV})\frac{1+\frac{\min(1-a^2,b^2)\min(a^2, 1-b^2)}{ab\sqrt{(1-a^2)(1-b^2)}}}{2}\label{cs3}
\end{align}
Combining (\ref{cs1}) and (\ref{cs3}), we have
\begin{equation}
-\lambda_{2}(\tilde{P}_{UV})\frac{1+\frac{\min(1-a^2,b^2)\min(a^2,1-b^2)}{ab\sqrt{(1-a^2)(1-b^2)}}}{2}
\le
\lambda\le
\lambda_{2}(\tilde{P}_{UV})\frac{1+\frac{\min(a^2,b^2)\min(1-a^2,1-b^2)}{ab\sqrt{(1-a^2)(1-b^2)}}}{2}
\end{equation}
$\blacksquare$
\bibliographystyle{unsrt}
\bibliography{refphd}
\end{document}